\documentclass[aps,pra,twocolumn,amssymb,amsfonts,floatfix,superscriptaddress,longbibliography,notitlepage,nofootinbib]{revtex4-1}
\usepackage[utf8]{inputenc}
\pdfoutput=1
\usepackage{color}
\usepackage{float}
\usepackage{xcolor}
\usepackage{mathtools,amsmath,amsthm,amsfonts,amssymb}
\usepackage{commath}
\usepackage{physics}
\usepackage{bm}
\usepackage[pdftex]{graphicx}
\usepackage[colorlinks=false]{hyperref}
\usepackage{enumitem}

\begin{document}

\title{Loss of integrability in a system with two-photon interactions}
\author{Fabrizio Ram\'irez}
\affiliation{Instituto de Ciencias Nucleares, Universidad Nacional Aut\'onoma de M\'exico, Apdo. Postal 70-543, C.P. 04510 Mexico City, Mexico}
\author{David Villase\~nor}
\email{d.v.pcf.cu@gmail.com}
\affiliation{CAMTP - Center for Applied Mathematics and Theoretical Physics, University of Maribor, Mladinska 3, SI-2000 Maribor, Slovenia, European Union}
\author{Viani S. Morales-Guzm\'an}
\author{Darly Y. Castro}
\author{Jorge G. Hirsch}
\email{hirsch@nucleares.unam.mx}
\affiliation{Instituto de Ciencias Nucleares, Universidad Nacional Aut\'onoma de M\'exico, Apdo. Postal 70-543, C.P. 04510 Mexico City, Mexico}

\begin{abstract}
Light-matter systems that exhibit two-photon interactions have emerged as powerful platforms for exploring quantum applications. In this work, we focus on the two-photon Dicke model, a system of significant experimental relevance that displays spectral collapse and undergoes a phase transition from a normal to a superradiant phase. We analyze the normal phase, where a classical limit with two degrees of freedom can be derived using a mean-field approximation. Our study presents a detailed investigation of the loss of integrability in the two-photon Dicke model, employing both quantum and classical diagnostics. These results allow us to explore various dynamical features of the system, including the onset of chaos and the existence of mixed phase-space behavior.
\end{abstract}

\maketitle

\section{Introduction}
\label{sec:Introduction}

Quantum systems involving the interaction between light and matter play a crucial role in a wide range of quantum technologies~\cite{Dowling2003}, including quantum information and communication~\cite{Zeilinger1999,Gisin2007,Swingle2016,Landsman2019}, quantum metrology and sensing~\cite{Taylor2016,Degen2017,Fiderer2018,Crawford2021,Montenegro2025}, and quantum simulation~\cite{Lloyd1996,Georgescu2014}, among others. The fundamental theoretical framework for describing such interactions is the Rabi model~\cite{Rabi1936,Rabi1937,Braak2016}, which considers a single two-level atom or qubit coupled to a single mode of a quantized electromagnetic field, allowing for the creation or annihilation of a single photon at a time. Extensions of the Rabi model to multiqubit systems have led to the Dicke model~\cite{Dicke1954,Kirton2019,Roses2020,Villasenor2024ARXIV}.

Integrable versions of the Rabi and Dicke models arise when counter-rotating terms in the Hamiltonian are neglected under the rotating wave approximation (RWA)~\cite{Scully2008Book}. This simplification yields the Jaynes-Cummings~\cite{Jaynes1963,Larson2021Book} and Tavis–Cummings models~\cite{Tavis1968,Garraway2011}, which correspond to the integrable single-qubit and multiqubit cases, respectively.

Introducing two-photon interactions into the Rabi model allows for the simultaneous creation or annihilation of two photons. Early studies of the two-photon Rabi model, particularly under the RWA, referred to as the two-photon Jaynes–Cummings model, investigated various quantum dynamical phenomena, including Rabi oscillations~\cite{Agarwal1985,Alsing1987}, squeezing~\cite{Puri1988,Toor1992}, and dissipative effects~\cite{Puri1988PRA}. More recent work has focused on the full two-photon Rabi model, exploring its exact solutions~\cite{Ng1999,Emary2002JPA,Emary2002,Dolya2008,Dolya2009}. The demonstration of integrability in the single-photon Rabi model~\cite{Braak2011} revitalized interest in the solvability of the two-photon counterpart~\cite{Chen2012,Travenec2012,Maciejewski2015,Travenec2015}, with several studies revealing a striking spectral collapse in the energy spectrum~\cite{Felicetti2015,Duan2016,Rico2020}, where discrete energy levels coalesce into a continuous band.

Subsequent investigations have further examined solvability aspects of the two-photon Rabi model and its variants~\cite{Peng2012,Peng2013,Maciejewski2017,Cui2017}, as well as quantum dynamical properties~\cite{Lu2017}. Although generalizations to multiphoton interactions have been proposed, it has been argued that processes involving more than two photons are unphysical in the Rabi model~\cite{Lo1998,Braak2026}. Nevertheless, recent findings highlight the potential of the two-photon Rabi model for quantum metrology applications~\cite{Ying2025AQT,Ying2025} and outline efficient schemes for generating two-photon processes in the presence of dissipation~\cite{Mazhari2025Arxiv}.

When extended to multiqubit systems, the two-photon Rabi model leads to the two-photon Dicke model, which also exhibits spectral collapse~\cite{Felicetti2015,Lo2021}. Research has investigated phenomena such as photon squeezing in this model~\cite{Banerjee2022} and its integrable limit~\cite{Gerry1989}, known as the two-photon Tavis–Cummings model. In the single-photon Dicke model, both ground-state and excited-state quantum phase transitions are well known~\cite{Emary2003,Emary2003PRL,Lambert2004,Brandes2005,Brandes2013}, giving rise to a superradiant phase marked by a significant increase in the average photon number. A similar collective superradiant phase has been identified in the two-photon Dicke model in the ultrastrong coupling regime~\cite{Garbe2017}.

Interestingly, multiphoton interactions beyond two photons are possible in the two-photon Dicke model due to the presence of special solutions known as dark states~\cite{Peng2017}, distinguishing it from the two-photon Rabi model, where such interactions are forbidden~\cite{Lo1998,Braak2026}. Further studies have also incorporated Stark shift terms into the model~\cite{Klimov1999,Zhai2025}. Experimental realizations of two-photon processes date back to early investigations of two-photon lasing~\cite{Loy1978,Schlemmer1980,Nikolaus1981,Gauthier1992} and masers~\cite{Brune1987,Brune1987PRL,Brune1988}, and have since been explored in systems such as quantum dots~\cite{Stufler2006,DelValle2010} and Rydberg atoms~\cite{Bertet2002,Zhang2013}. More recently, two-photon processes have been proposed for the development of quantum batteries~\cite{Crescente2020}.

Dissipative effects have also been studied within the two-photon Dicke model, particularly in relation to phase transitions~\cite{Garbe2020,Shah2025} and chaotic dynamics~\cite{Li2022,Li2024}. While progress has been made in understanding dissipative chaos, less attention has been given to the emergence of chaos in the isolated (nondissipative) two-photon Dicke model~\cite{Wang2019TPD}. In this work, we investigate the loss of integrability in the isolated two-photon Dicke model and characterize the onset of chaos in its normal phase using both classical and quantum diagnostics. Our results provide new insights into the nature of integrable, mixed, and fully chaotic regimes in this system. 

The remainder of this article is organized as follows: In Sec.~\ref{sec:DickeModel}, we introduce the two-photon Dicke model, discuss its general properties, and derive its classical limit in the normal phase via a mean-field approximation with coherent states. Section~\ref{sec:Integrability} is devoted to analyzing the integrable regime of the model, where we obtain analytical expressions for the eigenvalues and eigenstates using a Bogoliubov transformation, which are employed to characterize the loss of integrability in the low-energy regime. In Sec.~\ref{sec:IntegrabilityChaos}, we investigate the breakdown of integrability and characterize the onset of chaos in the normal phase through both classical and quantum indicators. Finally, in Sec.~\ref{sec:Conclusions}, we summarize our main results and present our conclusions. Additional technical details and derivations are provided in the appendices.

\section{Two-Photon Dicke model}
\label{sec:DickeModel}

The two-photon Dicke model describes the interaction between light and a set of $\mathcal{N}$ two-level atoms, resulting in the creation or annihilation of two bosonic excitations for each collective atomic excitation. The Hamiltonian, setting $\hbar=1$, is given by 
\begin{equation}
    \label{eq:DickeHamiltonian}
    \hat{H}_{\text{D}} = \omega \hat{a}^{\dagger} \hat{a} + \omega_{0} \hat{J}_{z} + \frac{\gamma}{\mathcal{N}} \left( \hat{a}^{\dagger 2} + \hat{a}^{2} \right)\left( \hat{J}_{+} + \hat{J}_{-} \right) ,
\end{equation}
where the operators $\hat{a}^\dagger$ and $\hat{a}$ are the creation and annihilation operators of the bosonic mode which satisfy the commutation relation $[\hat{a}, \hat{a}^{\dagger}] = \hat{\mathbb{I}}$. The operators $\hat{J}_{z}$ and $\hat{J}_{\pm}=\hat{J}_{x}\pm i\hat{J}_{y}$ are collective pseudospin operators, $\hat{J}_{x,y,z} = \frac{1}{2}\sum_{k = 1}^{\mathcal{N}}\hat{\sigma}_{x,y,z}^{k}$, that obey the SU(2) algebra of the Pauli matrices $\hat{\sigma}_{x,y,z}$. The parameters $\omega$ and $\omega_{0}$ represent the field frequency and the atomic level splitting, respectively. The parameter $\gamma$ is the atom-photon coupling strength, whose critical value $\gamma_{\text{sc}}=\omega/2$ identifies the spectral collapse~\cite{Felicetti2015,Duan2016,Rico2020}. 

The two-photon Dicke Hamiltonian commutes with the parity operator $\hat{\Pi} = e^{i\pi\hat{\Lambda}}$, $[\hat{H}_{\text{D}},\hat{\Pi}]=0$, where the operator $\hat{\Lambda}=\hat{a}^{\dagger}\hat{a}/2 + \hat{J}_{z} + j\hat{\mathbb{I}}$
defines the generalized number of excitations. This parity symmetry corresponds to the simultaneous exchange of the operators $\hat{a} \to i \hat{a}, \hat{a}^{\dagger} \to -i \hat{a}^{\dagger}$, and $\hat{J}_x \to -\hat{J}_x$ and divides the system into four subspaces identified by the eigenvalues of the parity operator $p=\pm 1,\pm i$. The Hamiltonian also commutes with the squared total pseudospin operator, $[\hat{H}_{\text{D}},\hat{\mathbf{J}}^{2}]=0$, whose eigenvalues $j(j+1)$ specify different subspaces. In this work, we study the case with $j=\mathcal{N}/2$.

The ground state of the two-photon Dicke model undergoes a phase transition from a normal phase, characterized by the absence of photons and atomic excitations, to a superradiant phase, in which both quantities attain finite values~\cite{Garbe2017,Banerjee2022}. The superradiant ground state is well described by a squeezed state. The critical value of the coupling strength that separates both phases is $\gamma_{\text{c}}=\sqrt{\omega\omega_0 j/2}$ ~\cite{Garbe2017}. The results presented in this work pertain to the normal phase ($\gamma<\gamma_{\text{c}}$) and consider coupling strength values below the spectral collapse threshold ($\gamma<\omega/2$).

\subsection{Classical two-photon Dicke Hamiltonian in the normal phase}
\label{sec:ClassicalDickeModel}

Previous research has demonstrated that squeezed states effectively describe the ground-state energy of the two-photon Dicke Hamiltonian in the normal and superradiant phase when using a mean-field approximation~\cite{Garbe2017}. As mentioned above, there is squeezing only in the superradiant phase. 

In this work, we apply a mean-field approximation using coherent states to derive the classical limit of the two-photon Dicke model in the normal phase. It accurately provides the ground-state energy and describes the complete classical dynamics in the normal phase of the system without dependency on the system size. The mean-field approximation follows the standard procedure established for the single-photon Dicke model~\cite{DeAguiar1992,Bakemeier2013,Bastarrachea2014a,Bastarrachea2014b,Bastarrachea2015,Chavez2016}. The complete derivation is presented in Appendix~\ref{app:MeanField}. We use Glauber coherent states
\begin{equation}
    \label{eq:GlauberState}
    |\alpha\rangle = e^{-|\alpha|^{2}/2}e^{\alpha\hat{a}^{\dagger}}|0\rangle 
\end{equation}
associated with the bosonic subspace and Bloch coherent states
\begin{equation}
    \label{eq:BlochState}
    |\beta\rangle = \frac{1}{(1+|\beta|^{2})^{j}}e^{\beta\hat{J}_{+}}|j,-j\rangle 
\end{equation}
associated with the atomic subspace, to define an overall Glauber-Bloch coherent state as the tensor product
\begin{equation}
    \label{eq:GlauberBlochState}
    |\mathbf{x}\rangle \equiv |\alpha\rangle\otimes|\beta\rangle,
\end{equation}
where $\mathbf{x}=(q,p;Q,P)$ are canonical position and momentum classical variables associated with the bosonic and atomic phase space through the parameters $\alpha(q,p)$ and $\beta(Q,P)$, respectively. The last parameters are defined in Eqs.~\eqref{eq:qp1} and~\eqref{eq:qp2}.

Computing the expectation value of the Hamiltonian given in Eq.~\eqref{eq:DickeHamiltonian} for the Glauber-Bloch coherent state in Eq.~\eqref{eq:GlauberBlochState}, we derive the classical Hamiltonian of the two-photon Dicke model
\begin{align}
    \label{eq:ClassicalDickeHamiltonian}
    h_{\text{D}}(\mathbf{x}) = & \frac{1}{j}\langle\mathbf{x}|\hat{H}_{\text{D}}|\mathbf{x}\rangle \\
    = & \frac{\omega}{2}\left(q^{2}+p^{2}\right) + \frac{\omega_{0}}{2}\left(Q^{2}+P^{2}\right)-\omega_{0} \nonumber \\
    & + \gamma\left(q^{2}-p^{2}\right)Q\sqrt{1-\frac{Q^{2}+P^{2}}{4}} , \nonumber 
\end{align}
appropriately scaled by the system size $j$. This scaling yields an intensive classical energy shell
\begin{equation}
    \epsilon \equiv \frac{E}{j}
\end{equation}
and defines an effective Planck constant $\hbar_{\text{eff}}=1/j$~\cite{Ribeiro2006}, which facilitates the semiclassical analysis.

\section{Integrability of the two-photon Dicke Hamiltonian}
\label{sec:Integrability}

When any of the three Hamiltonian parameters $\omega$, $\omega_0$, or $\gamma$ is set to zero, the two-photon Dicke model becomes integrable. Of particular interest is the case when the atomic level splitting $\omega_0$ is zero. This limit was analyzed in detail for the two-photon Rabi model in Ref.~\cite{Emary2002}. In this case, the system described by the two-photon Dicke model is in the superradiant phase~\cite{Garbe2017}.

\begin{figure*}[ht]
\centering
\includegraphics[width=1\textwidth]{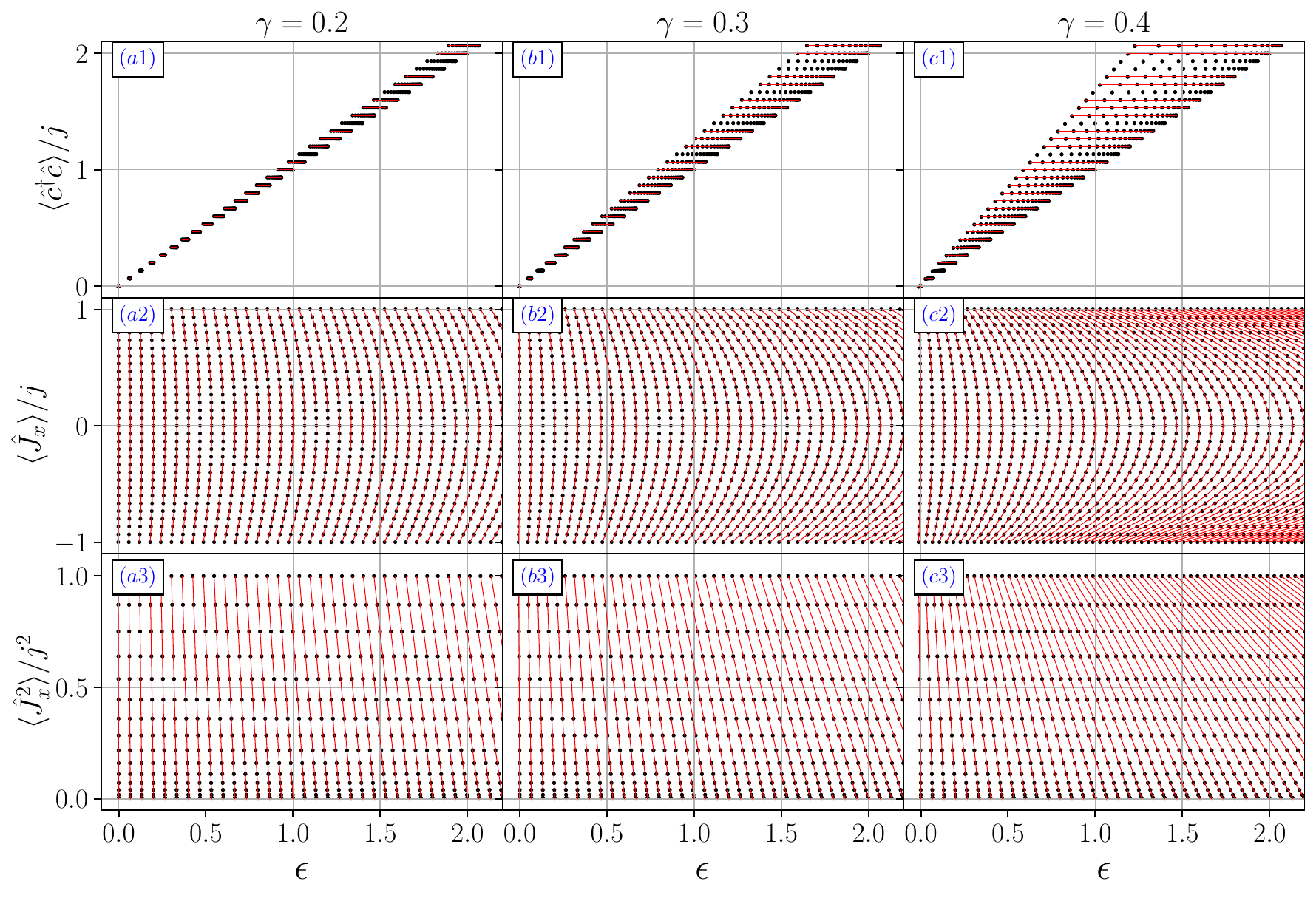}
\caption{Peres lattices of the operators (a1)-(c1) $\hat{c}^{\dagger}\hat{c}$, (a2)-(c2) $\hat{J}_{x}$, and (a3)-(c3) $\hat{J}_{x}^{2}$, as a function of the scaled energy $\epsilon=E/j$. Each column identifies a different coupling strength: (a1)-(a3) $\gamma=0.2$, (b1)-(b3) $\gamma=0.3$, and (c1)-(c3) $\gamma=0.4$. In each panel, the black dots represent the numerical expectation values using the eigenstates of the two-photon Dicke Hamiltonian in Eq.~\eqref{eq:DickeHamiltonian} and the red lines describe the analytical values using Eq.~\eqref{eq:AnalyticalSpectrum}. In panels (a1)-(c1) and (a2)-(c2), each red line represents $m_{x}$ as a continuous variable for discrete values of $n_{c}$. In panels (a3)-(c3), the continuous variable is $m_{x}^{2}$. System parameters: $\omega=1$, $\omega_{0} = 0$, and $j = 15$. We use a perturbation parameter $\varepsilon=0.001$. The truncation value is $n_{\text{max}} = 200$.}
\label{fig:PeresLatticeIntegrable}
\end{figure*}

\subsection{Integrable limit of zero atomic level splitting}
\label{sec:IntegrableLimit}

As mentioned above, by setting $\omega_0=0$ in Eq.~\eqref{eq:DickeHamiltonian}, the two-photon Dicke Hamiltonian becomes integrable. Employing a Bogoliubov transformation, as shown in the Appendix~\ref{app:Bogoliubov}, in this limit the two-photon Dicke Hamiltonian can be expressed as a set of harmonic oscillators, one for each eigenvalue $m_x$ of the operator $\hat{J}_x$,
\begin{equation}
    \label{eq:Hanalytic}
    \hat{H}'_{m_x} = \Omega_{m_x} \left(\hat{c}^\dagger \hat{c} + \frac{1}{2}\right) - \frac{\omega}{2} ,
\end{equation}
where $\Omega_{m_x} \equiv \omega\left(1 - 4 \lambda_{m_x}^2\right)^{1/2}$ and $\lambda_{m_x} \equiv 2\gamma m_x / (\omega\mathcal{N})$. The operators $\hat{c}$ and $\hat{c}^{\dagger}$ satisfy the commutation relation $[\hat{c},\hat{c}^{\dag}]=\hat{\mathbb{I}}$ and are defined through the modified annihilation operator
\begin{equation}
    \hat{c} = \frac{\hat{a} - \sigma_{m_x} \hat{a}^\dagger}{\sqrt{1 - \sigma_{m_x}^2}},
\end{equation}
where $\sigma_{m_x} = (\Omega_{m_x}/\omega - 1)/(2 \lambda_{m_x})$.

For the Hamiltonian in Eq.~\eqref{eq:Hanalytic}, the spectral collapse occurs when $\Omega_{m_x}=0$, associated with $\lambda_{m_x}=1/2$. For each value of $m_x$, the largest value that the coupling constant can have is
\begin{equation}
    \gamma_{\text{sc-x}} = \frac{\omega \mathcal{N}}{4 \left| m_x \right|} = \frac{j \omega}{2 \left| m_x \right|}.
\end{equation}
The smallest value $\gamma_{\text{sc}}= \omega/2$ occurs in the subspace with $m_x = \pm j$. This is the largest value that the coupling strength can have to avoid the spectral collapse. It makes a fundamental difference with the single-photon Dicke model, where the coupling strength $\gamma$ has no upper limit. 

The Hamiltonian in Eq.~\eqref{eq:Hanalytic} is diagonal in the basis
\begin{equation}
    \label{eq:ZeroSplittingBasis}
    |n_c;j,m_x\rangle = |n_{c}\rangle\otimes|j,m_x\rangle,
\end{equation} 
which identifies the eigenstates of the system. This basis satisfies the eigenvalue equations $\hat{c}^{\dagger} \hat{c} |n_c; j,m_x\rangle = n_c |n_c; j,m_x\rangle$ and $\hat{J}_x |n_c; j,m_x\rangle = m_x |n_c; j,m_x\rangle$, and the corresponding eigenenergies can be analytically expressed as
\begin{equation}
    \label{eq:AnalyticalSpectrum}
    E_{n_{c},m_{x}} = \Omega_{m_x} \left(n_{c} + \frac{1}{2}\right) - \frac{\omega}{2}.
\end{equation}

\subsection{Integrable correspondence}
\label{sec:Integrable}

It is instructive to compare the analytic expressions given above
with the numerical diagonalization of the two-photon Dicke Hamiltonian in Eq.~\eqref{eq:DickeHamiltonian}, when we set $\omega_{0}=0$. As explained in Appendix~\ref{app:Diagonalization}, in the general case,
the eigenvalue equation $\hat{H}_{\text{D}}|E_{k}\rangle = E_{k}|E_{k}\rangle$ 
is solved in the basis $|n;j,m_z\rangle = |n\rangle\otimes|j,m_z\rangle$. The numerical eigenstates $|E_{k}\rangle$ are used to evaluate the following expectation values 
\begin{align}
    \label{eq:Expectation_nc}
    n_c = & \langle E_{k}|\hat{c}^\dagger \hat{c}|E_{k}\rangle , \\
    \label{eq:Expectation_mx}
    m_x = & \langle E_{k}|\hat{J}_x|E_{k}\rangle , \\
    \label{eq:Expectation_mx2}
    m_x^2 = & \langle E_{k}|\hat{J}_x^2|E_{k}\rangle ,
\end{align}
which obviously
coincide with the analytical values given by Eq.~\eqref{eq:AnalyticalSpectrum}, when setting $\omega_{0}=0$.

A useful method for visually identifying the loss of integrability in a quantum system is the analysis of the energy-dependent expectation value of a chosen operator, $\langle E_{k}|\hat{O}|E_{k}\rangle$, a technique known as the Peres lattice~\cite{Peres1984PRL,Feingold1985,Feingold1986}. In this work, we compute the Peres lattices for the operators $\hat{c}^\dagger \hat{c}$, $\hat{J}_x$, and $\hat{J}_x^2$, as a function of the scaled eigenenergies $\epsilon_{k}=E_{k}/j$. As the pseudospin operator $\hat{J}_x$ does not commute with the system's parity operator, i.e., $[\hat{J}_x,\hat{\Pi}] \neq 0$,
its expectation value vanishes for eigenstates of definite parity due to orthogonality between states of different parity. To circumvent this and compute meaningful expectation values of $\hat{J}_x$, we explicitly break the parity symmetry by adding a small perturbative term to the Hamiltonian in Eq.~\eqref{eq:Htilde}, namely $\hat{H}'_{\text{D}} + \varepsilon \hat{J}_x$, where $\varepsilon$ is a small parameter controlling the strength of the perturbation.

In Fig.~\ref{fig:PeresLatticeIntegrable}, we plot the expectation values (black points) of the operators $\hat{c}^\dagger \hat{c}$, $\hat{J}_x$, and $\hat{J}_x^2$, which were numerically computed with Eqs.~\eqref{eq:Expectation_nc}-\eqref{eq:Expectation_mx2}, for a set of increasing couplings $\gamma$. In the same figure, we plot the analytical values (red lines) obtained with Eq.~\eqref{eq:AnalyticalSpectrum}. In Figs.~\ref{fig:PeresLatticeIntegrable}(a1)-\ref{fig:PeresLatticeIntegrable}(c1) and~\ref{fig:PeresLatticeIntegrable}(a2)-\ref{fig:PeresLatticeIntegrable}(c2), we display the value $m_{x}$ as a continuous variable (red lines) for discrete values of $n_{c}$. In Figs.~\ref{fig:PeresLatticeIntegrable}(a3)-\ref{fig:PeresLatticeIntegrable}(c3), the continuous variable is represented by $m_{x}^{2}$. As expected, in all panels, the numerical results are in exact agreement with the analytical predictions. 

Presenting the analytical expressions as continuous curves alongside the numerical data proves particularly useful for comparison, especially in the next section, where integrability is broken.

\begin{figure}[ht]
\centering
\includegraphics[width=1\columnwidth]{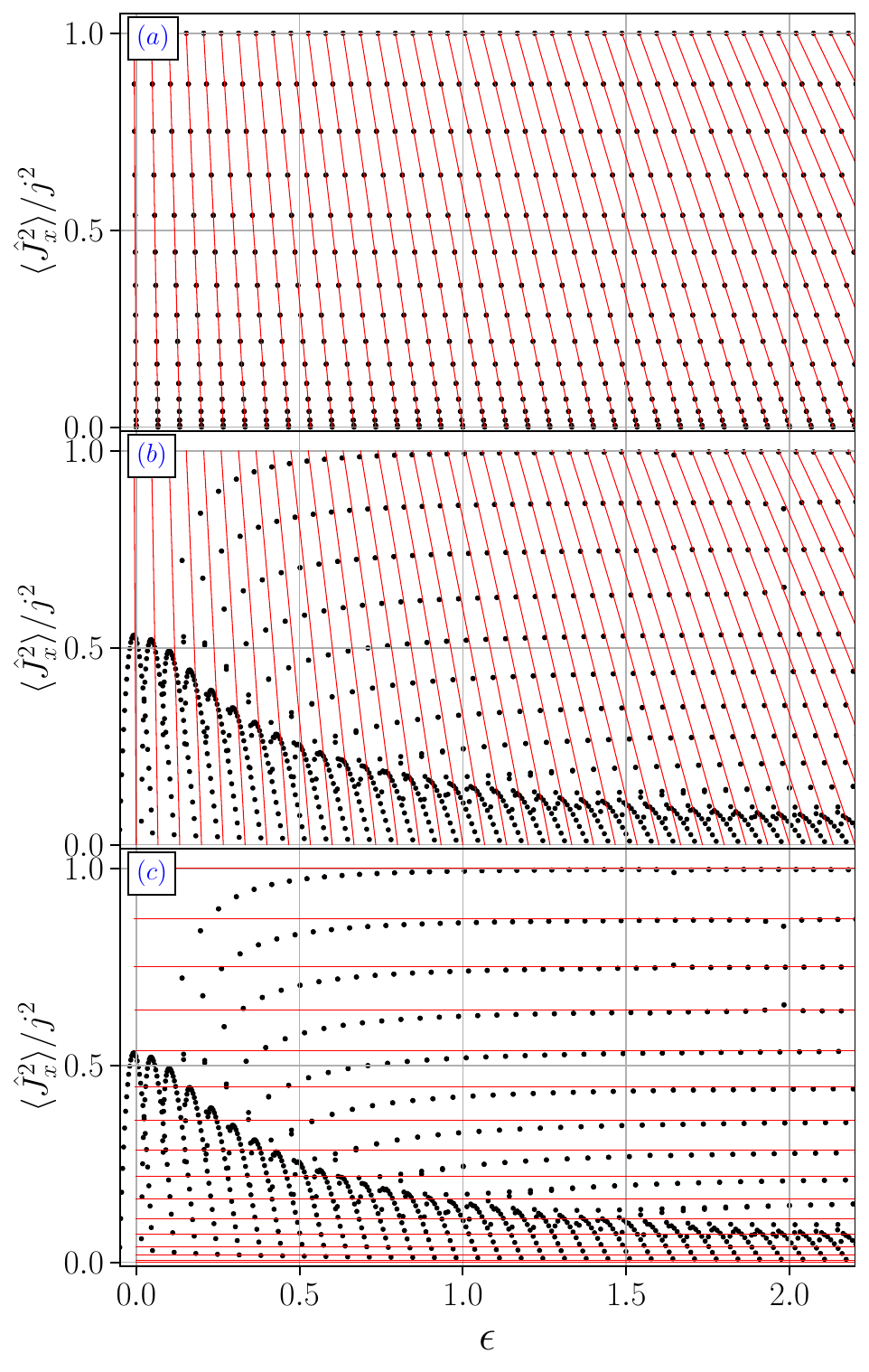}
\caption{Peres lattices of the operator $\hat{J}_{x}^{2}$ as a function of the scaled energy $\epsilon=E/j$. In each panel, the black dots represent the numerical expectation values using the eigenstates of the two-photon Dicke Hamiltonian in Eq.~\eqref{eq:DickeHamiltonian} and the red lines describe the analytical values using Eq.~\eqref{eq:AnalyticalSpectrum}. In panel (a), we set $\omega_0 = 0$ with $\varepsilon=0.001$ and plot $m_{x}^2$ as a continuous variable for discrete values of $n_{c}$. In panels (b) and (c), we set $\omega_{0} = 0.05$ to compute the numerical expectation values (black dots), but we use the analytical results (red lines) from panel (a). In panel (b), the continuous variable is $m_{x}^{2}$ as described in panel (a), while in panel (c), the continuous variable is $n_{c}$ for discrete values of $m_{x}^{2}$. System parameters: $\omega=1$, $\gamma=0.3$, and $j = 15$. The truncation value is $n_{\text{max}} = 200$.}
\label{fig:PeresLatticeNonIntegrable}
\end{figure}

\begin{figure}[ht]
    \centering
    \includegraphics[width=1\columnwidth]{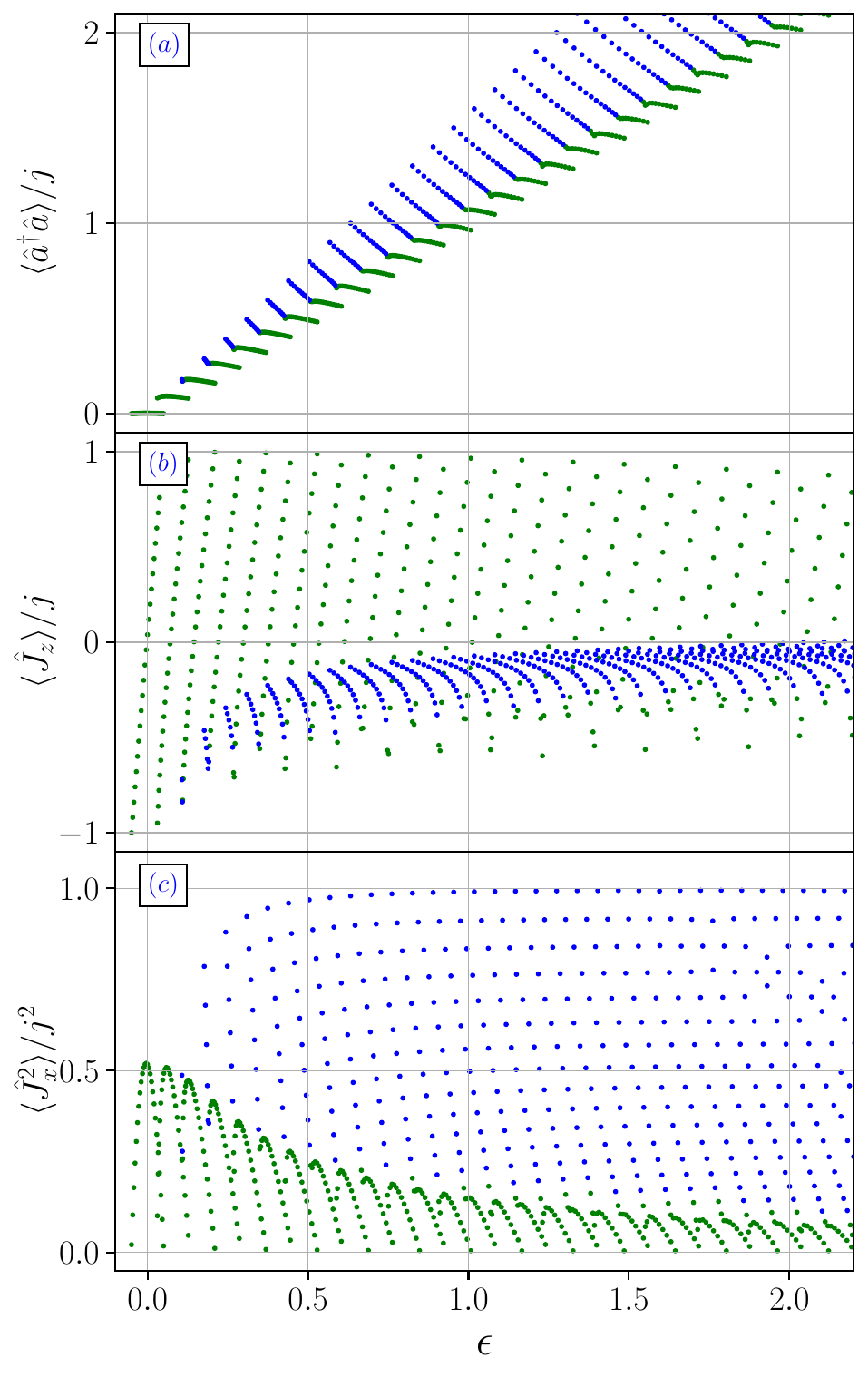}
    \caption{Peres lattices of the operators (a) $\hat{a}^{\dagger}\hat{a}$, (b) $\hat{J}_{z}$, and (c) $\hat{J}_{x}^{2}$, as a function of the scaled energy $\epsilon=E/j$. In each panel, the green and blue dots represent the numerical expectation values using the eigenstates of the two-photon Dicke Hamiltonian in Eq.~\eqref{eq:DickeHamiltonian}. The blue dots correspond to eigenstates that tend to behave as eigentates of the operator $\hat{J}_x$. The green dots correspond to eigenstates that tend to behave as eigenstates of the operator $\hat{J}_{z}$. System parameters: $\omega=1$, $\omega_0=0.05$, $\gamma=0.3$, and $j = 25$. The truncation value is $n_{\text{max}} = 200$.}
    \label{fig:PeresLatticeColor}
\end{figure}

\subsection{Breaking the integrability}
\label{sec:NonIntegrable}

We investigate the correspondence between the integrable case ($\omega_0 = 0$), described by Eq.~\eqref{eq:Hanalytic}, and the nonintegrable case ($\omega_0 \neq 0$), represented by Eq.~\eqref{eq:DickeHamiltonian}. As a first step, we focus on the effects of small atomic level splittings on the system's integrability and examine how deviations from the integrable limit influence the behavior of the observables.

Figure~\ref{fig:PeresLatticeNonIntegrable} presents the Peres lattice for the operator $\hat{J}_x^2$. The black dots correspond to numerical eigenvalues, while the red lines indicate the analytical predictions given by Eq.~\eqref{eq:AnalyticalSpectrum}. In Fig.~\ref{fig:PeresLatticeNonIntegrable}(a), we display the integrable case with $\omega_{0}=0$, where the red lines represent the continuous variable $m_{x}^{2}$ for discrete values of the photon number $n_{c}$. 

Introducing a finite atomic level splitting of $\omega_{0} = 0.05$ breaks the integrability, as illustrated in Fig.~\ref{fig:PeresLatticeNonIntegrable}(b). In this case, significant deviations between the numerical and analytical results appear, particularly in the lower part of the spectrum. Nevertheless, the analytical values from the integrable model still capture the general structure of the spectrum, and provide a reasonable approximation at higher energies. It can be seen in Fig.~\ref{fig:PeresLatticeNonIntegrable}(b) that for energies $\epsilon > 0.5$ the black dots are aligned with the red lines with different values of $n_c$, while in Fig.~\ref{fig:PeresLatticeNonIntegrable}(c) the black dots coincide with the horizontal lines with fixed values of $m_x^2$ for energies $\epsilon > 1$. 

To explore the observed regularities, in Fig.~\ref{fig:PeresLatticeColor}, we plot the Peres lattice of the operators $\hat{a}^{\dagger}\hat{a}$, $\hat{J}_{z}$, and $\hat{J}_{x}^{2}$. In Fig.~\ref{fig:PeresLatticeColor}(a), we highlight the expectation values that appear horizontally compressed in the Peres lattice of $\hat{a}^{\dagger}\hat{a}$, marking them in green. The remaining points, which are arranged along parallel diagonal lines, are shown in blue. This color scheme is then consistently applied to Fig.~\ref{fig:PeresLatticeColor}(b) and Fig.~\ref{fig:PeresLatticeColor}(c). The green points, originally associated with horizontally contracted structures in the Peres lattice of $\hat{a}^{\dagger}\hat{a}$, exhibit a dependence with the energy which is close to linear for $\hat{J}_z$, and parabolic for $\hat{J}_x^2$. On the other hand, the blue points, which were diagonally grouped in Fig.~\ref{fig:PeresLatticeColor}(a), show a linear dependence with the energy for $\hat{J}_x^2$, and a curved one for $\hat{J}_z$.

The last behaviors can be easily understood for the lowest-energy eigenstates. As shown in Fig.~\ref{fig:PeresLatticeColor}(a), the horizontal set of points, colored on green, have zero photons, and are very close to be eigenstates of $\hat{J}_z$, with energies $\omega_0 m_{z}, -j\leq m_{z}\leq j$, as can be seen in the lowest-energy region of Fig.~\ref{fig:PeresLatticeColor}(b). For these states, $\langle j,m_{z}|\hat{J}_{x}^{2}|j,m_{z}\rangle=[j(j+1)-m_{z}^{2}]/2$. The last expression describes the parabolic form observed in the first set of green points in Fig.~\ref{fig:PeresLatticeColor}(c), whose values run from $\langle \hat{J}_{x}^{2} \rangle/j^{2} = 1/(2j)$ for $m_{z}=\pm j$ to $\langle \hat{J}_{x}^{2} \rangle/j^{2} = (1+1/j)/2$ for $m_{z}=0$.

At higher energies, the eigenstates exhibit a competition between $\hat{J}_z$ and $\hat{J}_x$, with $\hat{J}_x$ tending to dominate when energy increases. This behavior can be observed at the horizontal arrangements of the blue dots in Fig.~\ref{fig:PeresLatticeColor}(c). This effect can be qualitatively understood by noticing that the expectation value of the photon number in Fig.~\ref{fig:PeresLatticeColor}(a) grows approximately linearly with energy.  This behavior is reasonable, since the contribution of the second term in the two-photon Dicke Hamiltonian to the scaled energy is at most $\omega_0$. Assuming that the expectation values $\langle \hat{a}^{\dagger 2} \rangle$ and $\langle \hat{a}^2 \rangle$ are of the same order of $\langle \hat{a}^{\dagger}\hat{a} \rangle$, the third term in the Hamiltonian scales as $\gamma \langle \hat{a}^{\dagger}\hat{a} \rangle$ and in the energy region $\epsilon > 1$ scales as $\gamma j$, which is significantly larger than $\omega_0$.

\section{Loss of integrability and onset of chaos}
\label{sec:IntegrabilityChaos}

The Peres lattices of various observables offer valuable insight into the integrability of the two-photon Dicke model and serve as effective tools for identifying the breakdown of integrability in the system. In general, the presence of well-ordered patterns in the lattice is indicative of integrable dynamics, while disordered or random-like structures signal a loss of integrability. This behavior reflects the nature of the classical limit of the quantum system, which can exhibit either regular or chaotic dynamics depending on the underlying parameters. In the following sections, we examine the connection between quantum and classical diagnostics in the two-photon Dicke model and explore how they reveal the emergence of chaos in the system.

\subsection{Classical and quantum chaos}
\label{sec:ClassicalQuantumChaos}

Classical integrability refers to a dynamical system that possesses a sufficient number of conserved quantities to constrain its motion to a lower-dimensional submanifold of the full phase space~\cite{OttBook}. A widely used qualitative tool for identifying the loss of integrability is the Poincar\'e section, which provides a visual representation of the system's dynamics. When the motion is predominantly regular, the Poincar\'e section displays well-organized, smooth curves. In contrast, chaotic dynamics gives rise to scattered, irregular, or random-like patterns. In many systems, a regime of mixed dynamics may arise, depending on specific parameter values, where both regular and chaotic behaviors coexist within the same phase space.

Quantum chaos refers to the study of spectral correlations in quantum systems whose classical counterparts exhibit chaotic behavior~\cite{Casati1980,Bohigas1984}. A widely used method for identifying such correlations is the analysis of nearest-neighbor level spacing distributions in the energy spectrum~\cite{HaakeBook,StockmannBook,WimbergerBook,Gutzwiller1990book}. In quantum systems that correspond to classically chaotic dynamics, the level spacing distribution typically follows that of the Gaussian Orthogonal Ensemble (GOE) from random matrix theory~\cite{Brody1981,Guhr1998,Fyodorov2011,MehtaBook}
\begin{equation}
    \label{eq:GOE}
        P_{\text{GOE}}(s) = \frac{\pi}{2}s\,e^{-\pi s^{2}/4}.
\end{equation}
The GOE distribution, often referred to as the Wigner-Dyson surmise, is characterized by level repulsion and a suppressed probability of small spacings. In the last expression, $s_{k}=(E_{k+1}-E_{k})/ S(E_{k})$ is the spacing between two neighboring energy levels $E_{k+1}>E_{k}$, scaled by the average spacing $S(E_{k})$. In contrast, the level spacing distribution of a regular (integrable) quantum system follows a Poisson distribution~\cite{Berry1977}
\begin{equation}
    \label{eq:Poisson}
        P_{\text{P}}(s) = e^{-s} ,
\end{equation}
reflecting the absence of level repulsion and indicating uncorrelated energy levels.

The need for careful treatment in the calculation of level spacing distributions, particularly regarding the unfolding of energy spectra to account for local density variations~\cite{Guhr1998}, has led to the development of alternative approaches for characterizing spectral correlations. One such method is the use of spectral ratios, which bypass the unfolding procedure and provide a robust indicator of level statistics. The spectral ratio is defined as follows~\cite{Oganesyan2007,Atas2013}
\begin{equation}
    \label{eq:SpectralRatio}
    r_{k} = \frac{\min(s_{k},s_{k-1})}{\max(s_{k},s_{k-1})} = \frac{\min(E_{k+1}-E_{k},E_{k}-E_{k-1})}{\max(E_{k+1}-E_{k},E_{k}-E_{k-1})} .
\end{equation}
By computing the mean value of the spectral ratio one can distinguish between chaotic and regular quantum systems. For spectra following the GOE distribution, the average ratio is $\langle r\rangle_{\text{GOE}}=4 - 2\sqrt{3} \approx 0.536$, whereas for Poisson-distributed spectra, the average is $\langle r\rangle_{\text{P}}=2\ln2 - 1\approx 0.386$~\cite{Oganesyan2007,Atas2013}. These characteristic values serve as effective indicators for identifying the presence of chaos or regularity in a given quantum system.

\begin{figure*}[ht]
\centering
\includegraphics[width=1\textwidth]{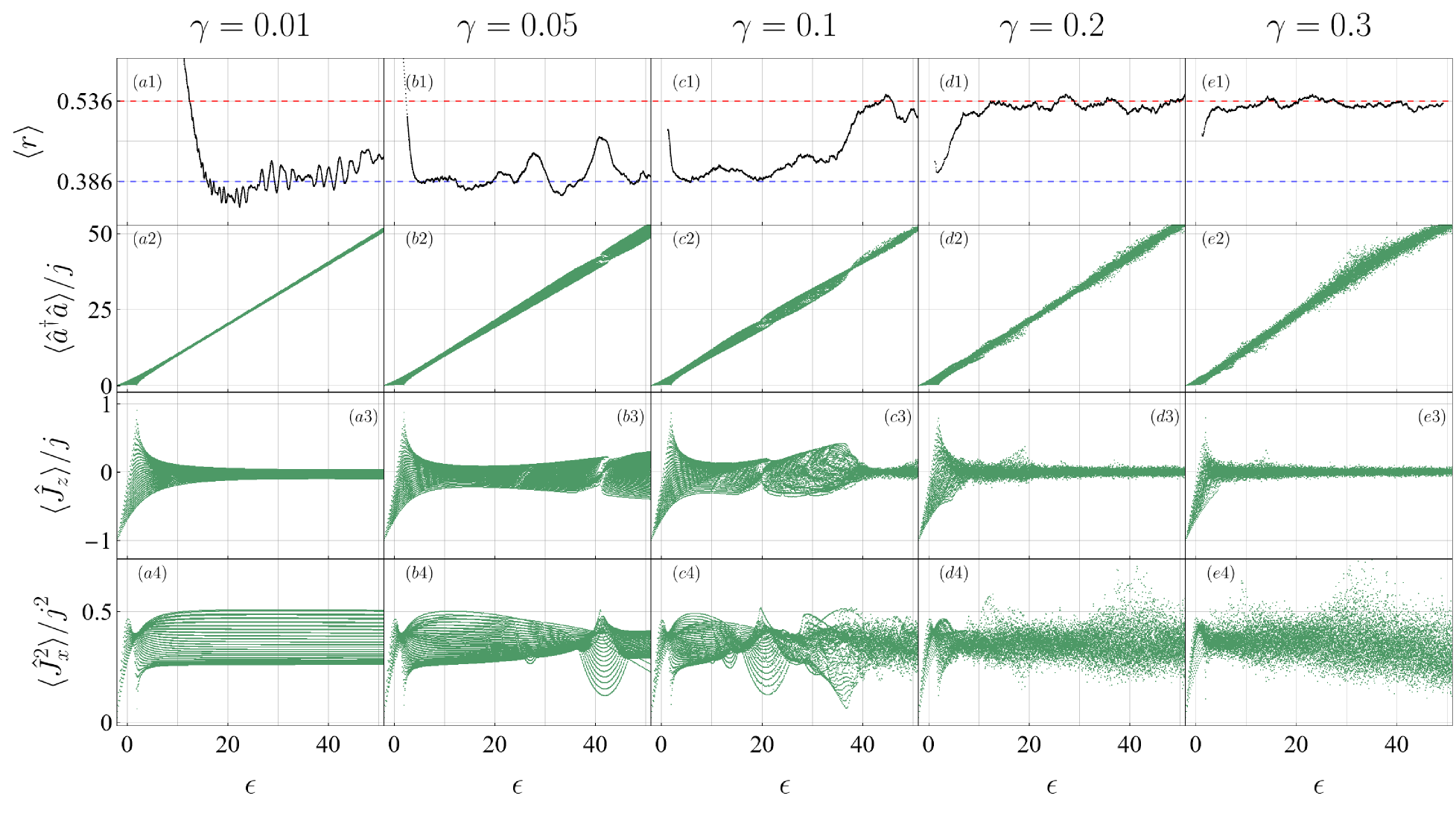}
\caption{(a1)-(e1) Average spectral ratio $\langle r \rangle$ and Peres lattices of the operators (a2)-(e2) $\hat{a}^{\dagger}\hat{a}$, (a3)-(e3) $\hat{J}_{z}$, and (a4)-(e4) $\hat{J}_{x}^{2}$ covering a wide energy spectrum, where $\epsilon=E/j$. Each column identifies a different coupling strength: (a1)-(a4) $\gamma=0.01$, (b1)-(b4) $\gamma=0.05$, (c1)-(c4) $\gamma=0.1$, (d1)-(d4) $\gamma=0.2$, and (e1)-(e4) $\gamma=0.3$. In panels (a1)-(e1), the black solid line represents the average spectral ratio in Eq.~\eqref{eq:SpectralRatio} using the converged eigenvalues of the two-photon Dicke Hamiltonian in Eq.~\eqref{eq:DickeHamiltonian}. We average the spectral ratio over the four parity subspaces to smooth spectral fluctuations. For each panel from (a1) to (e1), we get around $2\times10^4$ converged eigenvalues for each subspace. The red (blue) dashed line represents the analytical average spectral ratio of the GOE (Poisson) distribution. In panels (a2)-(e2), (a3)-(e3), and (a4)-(e4), the green dots represent the numerical expectation values using the eigenstates of the two-photon Dicke Hamiltonian for the parity subspace with eigenvalue $p=1$. System parameters: $\omega=1$, $\omega_0=2$, and $j = 25$. The truncation value is $n_{\text{max}} = 2\times10^3$.}
\label{fig:PeresLattice_omega2}
\end{figure*}

\begin{figure*}[ht]
\centering
\includegraphics[width=1\textwidth]{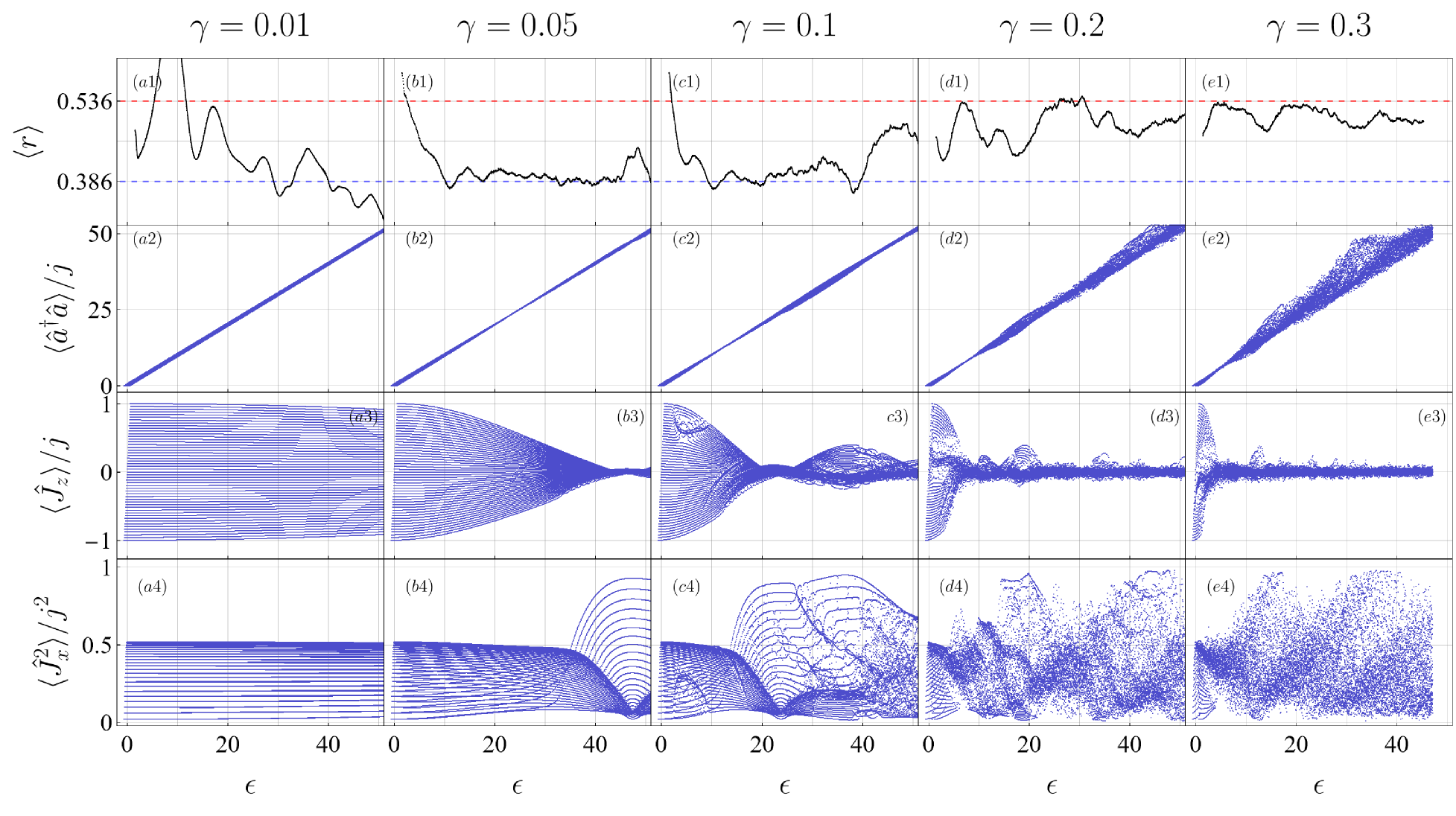}
\caption{(a1)-(e1) Average spectral ratio $\langle r \rangle$ and Peres lattices of the operators (a2)-(e2) $\hat{a}^{\dagger}\hat{a}$, (a3)-(e3) $\hat{J}_{z}$, and (a4)-(e4) $\hat{J}_{x}^{2}$ covering a wide energy spectrum, where $\epsilon=E/j$. Each column identifies a different coupling strength: (a1)-(a4) $\gamma=0.01$, (b1)-(b4) $\gamma=0.05$, (c1)-(c4) $\gamma=0.1$, (d1)-(d4) $\gamma=0.2$, and (e1)-(e4) $\gamma=0.3$. In panels (a1)-(e1), the black solid line represents the average spectral ratio in Eq.~\eqref{eq:SpectralRatio} using the converged eigenvalues of the two-photon Dicke Hamiltonian in Eq.~\eqref{eq:DickeHamiltonian}. We average the spectral ratio over the four parity subspaces to smooth spectral fluctuations. For each panel from (a1) to (e1), we get around $2\times10^4$ converged eigenvalues for each subspace. The red (blue) dashed line represents the analytical average spectral ratio of the GOE (Poisson) distribution. In panels (a2)-(e2), (a3)-(e3), and (a4)-(e4), the blue dots represent the numerical expectation values using the eigenstates of the two-photon Dicke Hamiltonian for the parity subspace with eigenvalue $p=1$. System parameters: $\omega=1$, $\omega_0=\pi/5$, and $j = 25$. The truncation value is $n_{\text{max}} = 2\times10^3$.}
\label{fig:PeresLattice_omega00p63}
\end{figure*}

\begin{figure*}[ht]
\centering
\includegraphics[width=1\textwidth]{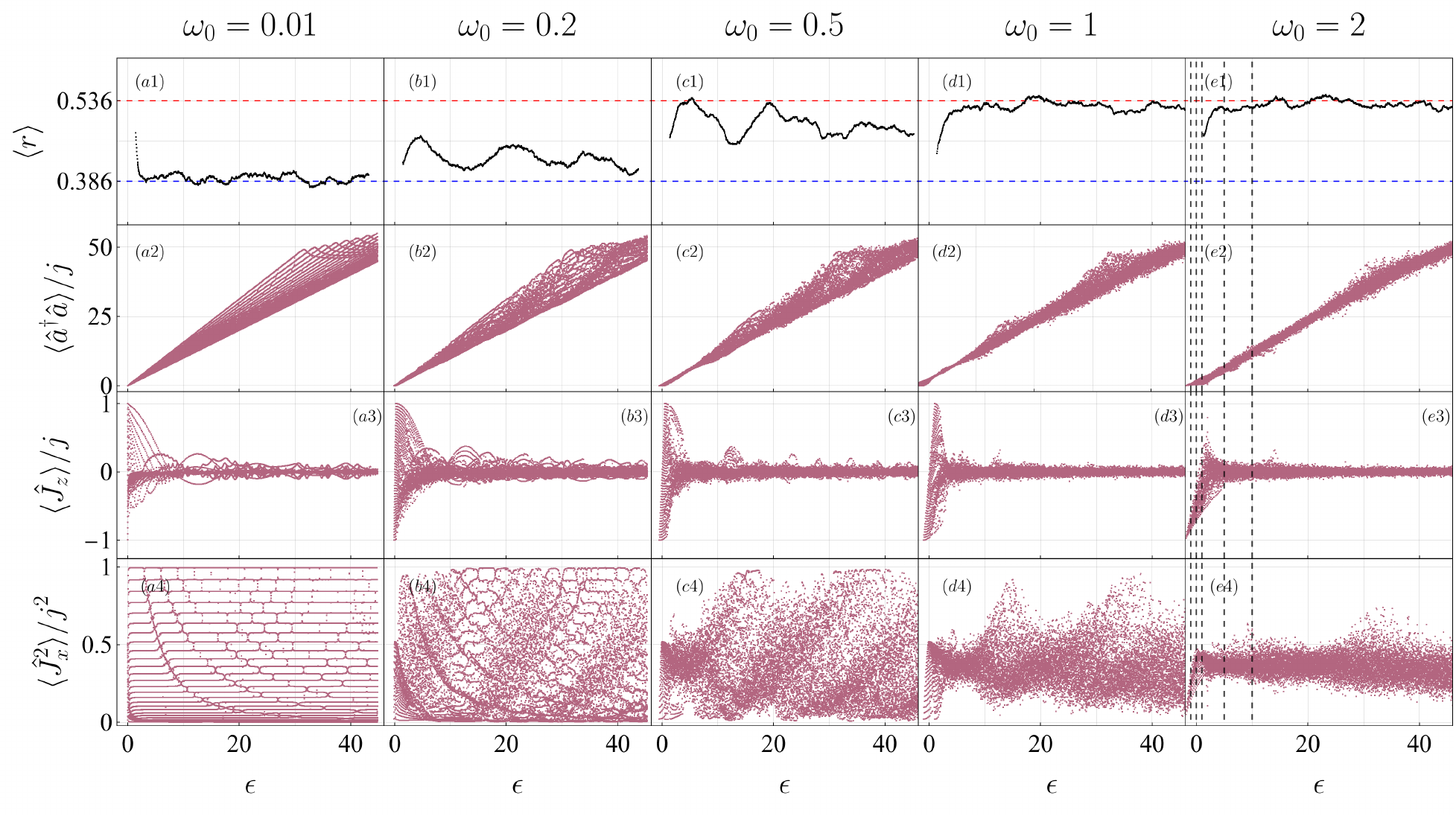}
\caption{(a1)-(e1) Average spectral ratio $\langle r \rangle$ and Peres lattices of the operators (a2)-(e2) $\hat{a}^{\dagger}\hat{a}$, (a3)-(e3) $\hat{J}_{z}$, and (a4)-(e4) $\hat{J}_{x}^{2}$ covering a wide energy spectrum, where $\epsilon=E/j$. Each column identifies a different atomic level splitting: (a1)-(a4) $\omega_{0}=0.01$, (b1)-(b4) $\omega_{0}=0.2$, (c1)-(c4) $\omega_{0}=0.5$, (d1)-(d4) $\omega_{0}=1$, and (e1)-(e4) $\omega_{0}=2$. In panels (a1)-(e1), the black solid line represents the average spectral ratio in Eq.~\eqref{eq:SpectralRatio} using the converged eigenvalues of the two-photon Dicke Hamiltonian in Eq.~\eqref{eq:DickeHamiltonian}. We average the spectral ratio over the four parity subspaces to smooth spectral fluctuations. For each panel from (a1) to (e1), we get around $2\times10^4$ converged eigenvalues for each subspace. The red (blue) dashed line represents the analytical average spectral ratio of the GOE (Poisson) distribution. In panels (a2)-(e2), (a3)-(e3), and (a4)-(e4), the pink dots represent the numerical expectation values using the eigenstates of the two-photon Dicke Hamiltonian for the parity subspace with eigenvalue $p=1$. In column (e1)-(e4), the vertical dashed lines represent the classical energy shells used in Fig.~\ref{fig:PoincareLyapunov}. System parameters: $\omega=1$, $\gamma=0.3$, and $j = 25$. The truncation value is $n_{\text{max}} = 2\times10^3$. }
\label{fig:PeresLattice_gamma0p3_S1}
\end{figure*}

\begin{figure*}[ht]
\centering
\includegraphics[width=1\textwidth]{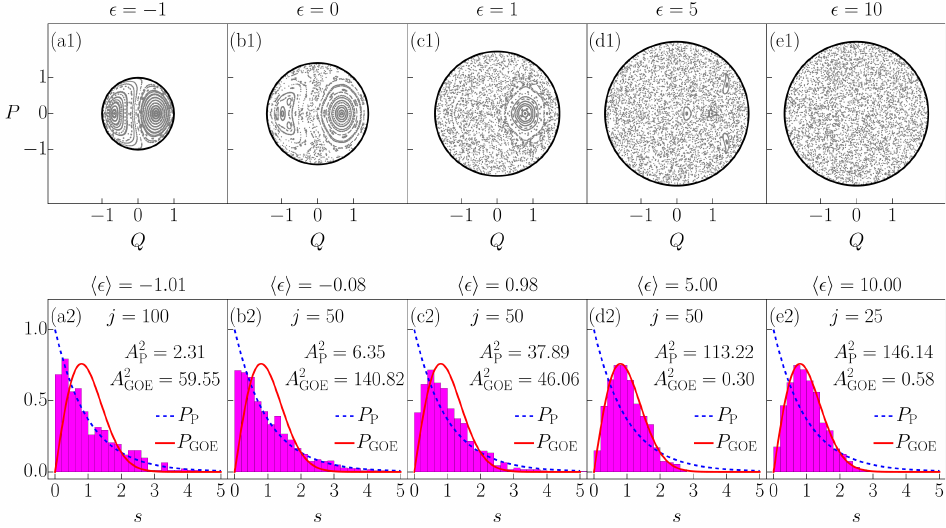}
\caption{(a1)-(e1) Poincar\'e sections projected on the atomic phase space of the classical two-photon Dicke Hamiltonian in Eq.~\eqref{eq:ClassicalDickeHamiltonian}. Each panel identifies a different classical energy shell: (a1) $\epsilon=-1$, (b1) $\epsilon=0$, (c1) $\epsilon=1$, (d1) $\epsilon=5$, and (e1) $\epsilon=10$. The black solid circle represents the available phase space of the selected energy shell. We choose initial conditions distributed over the phase space with $q_{0}=q_{+}(\epsilon,p_{0}=0,Q_{0},P_{0})$. The selected classical energy shells are shown in column (e1)-(e4) of Fig.~\ref{fig:PeresLattice_gamma0p3_S1} as vertical dashed lines. (a2)-(e2) Nearest-neighbor spacing distributions $P(s)$ for the converged eigenvalues of the two-photon Dicke Hamiltonian in Eq.~\eqref{eq:DickeHamiltonian}. In each panel, we use the converged eigenvalues contained in an energy window with mean $\langle \epsilon \rangle = \langle E \rangle/j$ and width $\sigma_{\epsilon}$: (a2) $\langle \epsilon \rangle = -1.01$ and $\sigma_{\epsilon} = 0.10$, (b2) $\langle \epsilon \rangle = -0.08$ and $\sigma_{\epsilon} = 0.55$, (c2) $\langle \epsilon \rangle = 0.98$ and $\sigma_{\epsilon} = 0.34$, (d2) $\langle \epsilon \rangle = 5.00$ and $\sigma_{\epsilon} = 0.26$, and (e2) $\langle \epsilon \rangle = 10.00$ and $\sigma_{\epsilon} = 1.35$. We use the parity subspace with eigenvalue $p=1$ and use different system sizes to get an appropriate number of eigenvalues inside the energy window: (a2) $j = 100$ and 461 eigenvalues, (b2)-(d2) $j=50$ and 1201 eigenvalues, and (e2) $j=25$ and 1601 eigenvalues. The red solid (blue dashed) line represents the analytical GOE (Poisson) distribution in Eq.~\eqref{eq:GOE} [Eq.~\eqref{eq:Poisson}]. Each panel shows the Anderson-Darling parameter, which ensures the agreement between the numerical data and the analytical distributions. System parameters: $\omega=1$, $\omega_{0}=2$, and $\gamma=0.3$. The truncation value is (a2) $n_{\text{max}} = 500$, (b2)-(d2) $n_{\text{max}} = 900$, and (e2) $n_{\text{max}} = 2\times10^{3}$.}
\label{fig:PoincareLyapunov}
\end{figure*}

\subsection{Integrability and coupling strength}
\label{sec:CouplingStrength}

In this section, we investigate the impact of the coupling strength 
$\gamma$ on the integrability of the system, focusing on values below the critical threshold at which spectral collapse occurs ($\gamma<\omega/2$). Our analysis is confined to the normal phase, characterized by a vanishing average photon number. We explore the breakdown of integrability across the energy spectrum, from the ground state to high-energy regions. To quantify the degree of spectral correlations, we employ the spectral ratio defined in Eq.~\eqref{eq:SpectralRatio}, averaging it over finite-energy windows to distinguish between GOE and Poisson statistics, thereby identifying signatures of quantum chaos or regular behavior in the system. We analyze two representative cases, one where the field frequency and the atomic level splitting are in resonance and one out of resonance.

\subsubsection{Resonant case: $\omega_0=2\omega$}

In Figs.~\ref{fig:PeresLattice_omega2}(a1)–\ref{fig:PeresLattice_omega2}(e1), we present the average spectral ratio $\langle r\rangle$ as a function of increasing coupling strength $\gamma$. To obtain these results, we diagonalize the two-photon Dicke Hamiltonian following the procedure outlined in Appendix~\ref{app:Diagonalization}, using an atomic level splitting in resonance with the field frequency, $\omega=1$ and $\omega_{0} = 2\omega = 2$. The energy spectra from all four parity subspaces are computed, and the spectral ratio is calculated within each subspace by averaging over finite-energy windows. To mitigate strong spectral fluctuations and enhance the robustness of the analysis, we subsequently average the $\langle r\rangle$ values in the four subspaces.

The remaining rows in Fig.~\ref{fig:PeresLattice_omega2} display the Peres lattices for various observables corresponding to the same set of increasing coupling strengths. Figures~\ref{fig:PeresLattice_omega2}(a2)–\ref{fig:PeresLattice_omega2}(e2) show the Peres lattices of the photon number operator $\hat{a}^{\dagger}\hat{a}$, while Figs.~\ref{fig:PeresLattice_omega2}(a3)–\ref{fig:PeresLattice_omega2}(e3) and~\ref{fig:PeresLattice_omega2}(a4)–\ref{fig:PeresLattice_omega2}(e4) display those of the pseudospin operators $\hat{J}_{z}$ and $\hat{J}_{x}^{2}$, respectively. For clarity in visualizing the structure of the Peres lattices, we restrict this part of the analysis to a single-parity subspace with eigenvalue $p=1$.

Overall, the coupling strength $\gamma$ plays a key role in driving a transition from the integrable to the nonintegrable regime of the system. This transition is clearly reflected in the increasing disorder of the Peres lattices as $\gamma$ increases. The loss of integrability is particularly evident in the Peres lattices of the operator $\hat{J}_{x}^{2}$ [see Figs.~\ref{fig:PeresLattice_omega2}(a4)–\ref{fig:PeresLattice_omega2}(e4)], where the initially ordered structures progressively give way to irregular, chaotic patterns. Notably, even in those lattices that exhibit significant disorder at higher energies, ordered patterns persist in the low-energy region [see Figs.~\ref{fig:PeresLattice_omega2}(d2)–\ref{fig:PeresLattice_omega2}(d4) and~\ref{fig:PeresLattice_omega2}(e2)–\ref{fig:PeresLattice_omega2}(e4), column view], indicating that the system retains integrable characteristics near the ground state.

In Figs.~\ref{fig:PeresLattice_omega2}(a1)–\ref{fig:PeresLattice_omega2}(c1), the average spectral ratio $\langle r \rangle$ shows a decay from large values at low energies, followed by fluctuations around the value associated with Poisson statistics. These elevated values of $\langle r \rangle$ are characteristic of nearly harmonic spectra, such as that of the harmonic oscillator, for which $\langle r \rangle=1$. Notably, in Fig.~\ref{fig:PeresLattice_omega2}(c1), the spectral ratio increases again at higher energies, approaching values characteristic of GOE statistics. This shift is consistent with the emergence of disorder in the corresponding Peres lattices at high energies [see Figs.~\ref{fig:PeresLattice_omega2}(c2)–\ref{fig:PeresLattice_omega2}(c4), column view], further indicating the onset of nonintegrable dynamics in that energy regime. At these relatively low values of $\gamma$, we observe peculiar sets of regular lines in the expectation values of $\hat{J}_{x}^{2}$ at high energies, $\epsilon \approx 40$ in Fig.~\ref{fig:PeresLattice_omega2}(b4) and $\epsilon \approx 20$ in Fig.~\ref{fig:PeresLattice_omega2}(c4). These unexpectedly ordered structures in the middle of an otherwise disordered regime suggest underlying mechanisms that warrant further investigation.

The final panels, Fig.~\ref{fig:PeresLattice_omega2}(d1) and~\ref{fig:PeresLattice_omega2}(e1) show that the average spectral ratio $\langle r\rangle$ stabilizes around the GOE limit at high energies, following a transition from Poisson-like statistics at lower energies. This behavior aligns with the corresponding Peres lattices, which display pronounced disorder in the high-energy regime and ordered structures at low energies [see Figs.~\ref{fig:PeresLattice_omega2}(d2)–\ref{fig:PeresLattice_omega2}(d4) and~\ref{fig:PeresLattice_omega2}(e2)–\ref{fig:PeresLattice_omega2}(e4), column view].

In general, for the resonant case with $\omega_0=2\omega$, we observe a clear transition from Poisson to GOE statistics and from ordered to disordered Peres lattices as the coupling strength $\gamma$ increases from low to high values. This agreement reinforces the complementary nature of these quantum chaos indicators and underscores the spectral ratio's usefulness as a diagnostic tool for integrability breakdown in the two-photon Dicke model.

\subsubsection{Nonresonant case: $\omega_0 = (\pi/5)\omega$}

We now extend the previous analysis to an atomic level splitting out of resonance with the field frequency, $\omega = 1$ and $\omega_{0} = (\pi/5)\omega \approx 0.628$. In Fig.~\ref{fig:PeresLattice_omega00p63}, we present the same set of diagnostics as before, the average spectral ratio 
$\langle r\rangle$ and the Peres lattices corresponding to the operators $\hat{a}^{\dagger}\hat{a}$, $\hat{J}_{z}$, and $\hat{J}_{x}^{2}$. These are shown in Figs.~\ref{fig:PeresLattice_omega00p63}(a1)–\ref{fig:PeresLattice_omega00p63}(e1), ~\ref{fig:PeresLattice_omega00p63}(a2)–\ref{fig:PeresLattice_omega00p63}(e2), ~\ref{fig:PeresLattice_omega00p63}(a3)–\ref{fig:PeresLattice_omega00p63}(e3), and~\ref{fig:PeresLattice_omega00p63}(a4)–\ref{fig:PeresLattice_omega00p63}(e4), respectively.

As in the resonant case with $\omega_0=2\omega$, we observe for the nonresonant case with $\omega_{0}=(\pi/5)\omega$ a consistent trend in which increasing the coupling strength $\gamma$ drives a transition from integrable to nonintegrable behavior. This transition is again clearly reflected in the increasing disorder of the Peres lattices, particularly for the operator $\hat{J}_{x}^{2}$ [see Figs.~\ref{fig:PeresLattice_omega00p63}(a4)–\ref{fig:PeresLattice_omega00p63}(e4)]. The persistence of integrable dynamics near the ground state is newly detected for the low-energy region of the Peres lattices that exhibit strong disorder at higher energies [see Figs.~\ref{fig:PeresLattice_omega00p63}(d2)–\ref{fig:PeresLattice_omega00p63}(d4) and~\ref{fig:PeresLattice_omega00p63}(e2)–\ref{fig:PeresLattice_omega00p63}(e4), column view].

The behavior of the average spectral ratio $\langle r\rangle$ for the nonresonant case with $\omega_0=(\pi/5)\omega$ also exhibits an overall trend similar to that observed in the resonant case with $\omega_0=2\omega$. A special case is first illustrated in Fig.~\ref{fig:PeresLattice_omega00p63}(a1) for the smallest coupling strength $\gamma = 0.01$, where significant deviations from the expected Poisson statistics are observed. Despite the corresponding Peres lattices displaying well-ordered structures [see Figs.~\ref{fig:PeresLattice_omega00p63}(a2)–\ref{fig:PeresLattice_omega00p63}(a4), column view], the spectral ratio fails to accurately reflect the integrable nature of the system.

As occurred in the resonant case [see Figs.~\ref{fig:PeresLattice_omega2}(a1)-\ref{fig:PeresLattice_omega2}(c1)], a similar behavior is observed for the nonresonant case in Fig.~\ref{fig:PeresLattice_omega00p63}(b1), where $\langle r \rangle$ initially takes on large values in the low-energy region and then decays toward values consistent with Poisson statistics at intermediate energies. This observation is consistent with the nearly equidistant lines representing the expectation values of $\hat{J}_z$ in Fig.~\ref{fig:PeresLattice_omega00p63}(a3), and in the low-energy region of Fig.~\ref{fig:PeresLattice_omega00p63}(b3). In Fig.~\ref{fig:PeresLattice_omega00p63}(b4), at energies $\epsilon > 35$ the energy eigenstates tend to be also eigenstates of the operator $\hat{J}_x$, opening along the different values of $m_x^{2}$, which also explains that the expectation values of $\hat{J}_z$ in Fig.~\ref{fig:PeresLattice_omega00p63}(b3) tend to zero.

A similar initial trend in $\langle r \rangle$ is observed in Fig.~\ref{fig:PeresLattice_omega00p63}(c1), where it has large values at low energies before converging toward the Poisson limit. However, at higher energies, $\langle r \rangle$ begins to rise again, signaling the onset of spectral correlations. This behavior aligns with the increasing disorder observed in the corresponding Peres lattices [see Figs.~\ref{fig:PeresLattice_omega00p63}(c2)–\ref{fig:PeresLattice_omega00p63}(c4), column view], marking the beginning of a transition toward nonintegrability.

Figures~\ref{fig:PeresLattice_omega00p63}(d1) and~\ref{fig:PeresLattice_omega00p63}(e1) show that the average spectral ratio $\langle r\rangle$ fluctuates around the value expected for GOE statistics in certain energy regions. This behavior is consistent with the strong disorder observed in the corresponding Peres lattices at high energies [see Figs.~\ref{fig:PeresLattice_omega00p63}(d2)–\ref{fig:PeresLattice_omega00p63}(d4) and~\ref{fig:PeresLattice_omega00p63}(e2)–\ref{fig:PeresLattice_omega00p63}(e4), column view].

With the exception of the case $\gamma = 0.01$, where the energy spectrum remains nearly harmonic and thus obscures the spectral statistics, we observe again a strong correspondence between the degree of disorder in the Peres lattices and the values of the average spectral ratio $\langle r\rangle$. As the coupling strength $\gamma$ increases, a clear trend emerges in which the spectral ratio transitions from values characteristic of Poisson statistics to those indicative of GOE statistics, reflecting a progressive loss of integrability and the emergence of quantum chaos.

Having similar behaviors in the route to chaos observed in the average spectral ratio $\langle r\rangle$, there are noticeably differences between the two cases analyzed above. For the nonresonant case with $\omega_0 = (\pi/5)\omega$, Figs.~\ref{fig:PeresLattice_omega00p63}(a3)-\ref{fig:PeresLattice_omega00p63}(c3) show a correspondence between the expectation values of $\hat{J}_z$ and the eigenvalues $m_z$, which is clearly absent for the resonant case with $\omega_0=2\omega$, shown in Figs.~\ref{fig:PeresLattice_omega2}(a3)-\ref{fig:PeresLattice_omega2}(c3).

\subsection{Integrability and atomic level splitting}
\label{sec:AtomicSplitting}

Complementing the analysis in the previous section, we now examine the effect of the atomic level splitting  $\omega_0$ on the integrability of the system. This study is conducted for coupling strengths below the critical threshold at which spectral collapse occurs ($\gamma < \omega/2$).

In Figs.~\ref{fig:PeresLattice_gamma0p3_S1}(a1)–\ref{fig:PeresLattice_gamma0p3_S1}(e1), we present the average spectral ratio $\langle r\rangle$ for increasing values of the atomic level splitting $\omega_0$, fixing the coupling strength at $\gamma = 0.3$. The figure follows the same organization used previously, showing $\langle r\rangle$ alongside the Peres lattices for the operators $\hat{a}^{\dagger}\hat{a}$, $\hat{J}_{z}$, and $\hat{J}_{x}^{2}$. Each indicator is presented in Figs.~\ref{fig:PeresLattice_gamma0p3_S1}(a1)–\ref{fig:PeresLattice_gamma0p3_S1}(e1), ~\ref{fig:PeresLattice_gamma0p3_S1}(a2)–\ref{fig:PeresLattice_gamma0p3_S1}(e2), ~\ref{fig:PeresLattice_gamma0p3_S1}(a3)–\ref{fig:PeresLattice_gamma0p3_S1}(e3), and~\ref{fig:PeresLattice_gamma0p3_S1}(a4)–\ref{fig:PeresLattice_gamma0p3_S1}(e4), respectively.

Similar to the trends observed in the previous section with increasing coupling strength $\gamma$, we find that as the atomic level splitting $\omega_0$ increases, the system transitions from the integrable to the nonintegrable regime. This is evident from the growing disorder in the Peres lattices as $\omega_0$ increases. The disorder is most prominent in the sequence of Peres lattices of the operator $\hat{J}_{x}^{2}$ [see Figs.~\ref{fig:PeresLattice_gamma0p3_S1}(a4)–\ref{fig:PeresLattice_gamma0p3_S1}(e4)]. However, even in cases with strong disorder, ordered patterns can still be identified in the low-energy regions, as shown in Figs.~\ref{fig:PeresLattice_gamma0p3_S1}(c2)–\ref{fig:PeresLattice_gamma0p3_S1}(c4), ~\ref{fig:PeresLattice_gamma0p3_S1}(d2)–\ref{fig:PeresLattice_gamma0p3_S1}(d4), and~\ref{fig:PeresLattice_gamma0p3_S1}(e2)–\ref{fig:PeresLattice_gamma0p3_S1}(e4) in column view.

The general trend of the average spectral ratio $\langle r\rangle$ clearly demonstrates the system's transition from Poisson statistics at low atomic level splittings to GOE statistics at higher values. This transition is well aligned with the increasing disorder observed in the Peres lattices. In Fig.~\ref{fig:PeresLattice_gamma0p3_S1}(b1), we observe a regime of mixed statistics, where $\langle r\rangle$ does not align with either Poisson or GOE statistics. The corresponding Peres lattices exhibit a mixed pattern of order and disorder [see Figs.~\ref{fig:PeresLattice_gamma0p3_S1}(b2)-\ref{fig:PeresLattice_gamma0p3_S1}(b4), column view]. This behavior could provide valuable insights into the characterization of mixed-type systems~\cite{Berry1984,Robnik1998,Robnik2023}.

\subsection{Onset of classical and quantum chaos}
\label{sec:Chaos}

In the previous sections, we demonstrated how the Peres lattices serve as powerful tools for visually detecting the loss of integrability in the two-photon Dicke model. The behavior of the system is primarily controlled by the modulation of two key parameters: the coupling strength $\gamma$ and the atomic level splitting $\omega_0$. These parameters can thus act as control variables for the system's integrability. Furthermore, we observed a general agreement between the order (disorder) displayed in the Peres lattices and the Poisson (GOE) statistics of the energy spectrum, as detected through the average spectral ratio $\langle r \rangle$.

In this section, we investigate the classical dynamics of the two-photon Dicke model using the classical Hamiltonian given in Eq.~\eqref{eq:ClassicalDickeHamiltonian}. To explore the system's behavior, we use Poincar\'e sections for various classical energy shells $\epsilon$, corresponding to regions where the quantum system exhibits either integrable or nonintegrable behavior, as seen in the Peres lattices of the given operators. The second-order equation $h_{\text{D}}(q,p;Q,P)=\epsilon$ yields two roots for the variable $q=q_{\pm}(\epsilon,p,Q,P)$, and we select the positive root to define the initial conditions in the phase space, specifically setting $q_{0}=q_{+}(\epsilon,p_{0}=0,Q_{0},P_{0})$.

In Figs.~\ref{fig:PoincareLyapunov}(a1)-\ref{fig:PoincareLyapunov}(e1), we present Poincar\'e sections for an ensemble of initial conditions evolved in phase space using the two-photon Dicke Hamiltonian in Eq.~\eqref{eq:ClassicalDickeHamiltonian}, for increasing values of the classical energy shell $\epsilon$. These Poincar\'e sections are projected onto the atomic phase space defined by the classical variables $(Q,P)$. We fix the coupling strength at $\gamma=0.3$ and use the resonant case of the atomic level splitting $\omega_{0}=2\omega$, which corresponds to the parameter sets shown in Figs.~\ref{fig:PeresLattice_omega2}(e1)-\ref{fig:PeresLattice_omega2}(e4) and Figs.~\ref{fig:PeresLattice_gamma0p3_S1}(e1)-\ref{fig:PeresLattice_gamma0p3_S1}(e4) in column view. The chosen classical energy shells are within the region where we observe ordered patterns in the Peres lattices, transitioning to disorder as the average spectral ratio shifts from Poisson to GOE statistics. The last energy shells are displayed as vertical dashed lines in Figs.~\ref{fig:PeresLattice_gamma0p3_S1}(e1)-\ref{fig:PeresLattice_gamma0p3_S1}(e4) in column view. As seen in Figs.~\ref{fig:PoincareLyapunov}(a1)-\ref{fig:PoincareLyapunov}(e1), we detect an overall transition from integrability to chaotic motion with increasing energy. Specifically, the toroidal shapes in the Poincar\'e sections start to contract, and random patterns gradually emerge, indicating a shift toward complete ergodicity.

As a final step, we compute the nearest-neighbor spacing distribution for a set of converged eigenvalues of the two-photon Dicke model, using the parity subspace with eigenvalue $p=1$. We focus on a set of eigenvalues centered around the classical energy shell corresponding to the Poincar\'e sections presented in Figs.~\ref{fig:PoincareLyapunov}(a1)-\ref{fig:PoincareLyapunov}(e1). To unfold these eigenvalues, we apply a local averaging procedure and present the spacing distributions in Figs.~\ref{fig:PoincareLyapunov}(a2)-\ref{fig:PoincareLyapunov}(e2). Due to the broad energy range, we utilize different system sizes to obtain an appropriate number of eigenvalues for computing the spectral statistics. Specifically, in Fig.~\ref{fig:PoincareLyapunov}(a2), we use a large system size of $j=100$ to capture enough energy levels in the lower part of the spectrum, while in Fig.~\ref{fig:PoincareLyapunov}(e2), we use a smaller system size $j=25$ to be able to reach the high-energy region. For intermediate energies, we use a system size of $j=50$ in Figs.~\ref{fig:PoincareLyapunov}(b2)-\ref{fig:PoincareLyapunov}(d2). In these plots, we observe an overall transition from Poisson to GOE statistics with increasing energy. To ensure the correspondence between the numerical data and the analytical distributions presented in Eq.~\eqref{eq:GOE} and Eq.~\eqref{eq:Poisson}, we perform the Anderson-Darling test~\cite{Anderson1952}. The values of the Anderson-Darling parameter, denoted $A^{2}$, are shown in each panel, where a value of $A^{2}\leq2.5$ confirms that the numerical data are well-described by the corresponding analytical distribution.

Regarding the quantum-classical correspondence, we observe a reasonable agreement between the spectral statistics and the classical motion identified in the Poincar\'e sections. In Figs.~\ref{fig:PoincareLyapunov}(a1) and~\ref{fig:PoincareLyapunov}(a2), we see the correspondence between integrable classical motion and Poisson statistics. In contrast, Figs.~\ref{fig:PoincareLyapunov}(e1) and~\ref{fig:PoincareLyapunov}(e2) show the correspondence between fully chaotic classical motion and GOE statistics. Additionally, in Fig.~\ref{fig:PoincareLyapunov}(d1), we observe small stability islands surrounded by a chaotic sea, which do not significantly affect the spectral statistics described by the GOE distribution shown in Fig.~\ref{fig:PoincareLyapunov}(d2). For the other cases, we identify mixed classical dynamics, with corresponding spectral statistics that transition between Poisson and GOE distributions.

\section{Conclusions}
\label{sec:Conclusions}

In this study, we carried out a comprehensive investigation into the loss of integrability in the two-photon Dicke model. We began by analyzing the integrable limit at zero atomic level splitting within the superradiant phase of the system, showing that the analytical solutions in this regime offer valuable insights for characterizing the breakdown of integrability in the low-energy spectrum. Building on this foundation, we shifted our focus to the normal phase, where we employed Peres lattices as a visual tool to track the onset of nonintegrable behavior.

We found that integrability is lost in the normal phase either by increasing the coupling strength while holding the atomic level splitting constant, or by increasing the atomic level splitting at fixed coupling. The average spectral ratio $\langle r \rangle$ exhibited strong agreement with the disorder patterns seen in the Peres lattices: energy regions with disordered lattices aligned with GOE statistics, while regions showing ordered structures conformed to Poisson statistics.

To explore the classical dynamics, we used a mean-field approximation with coherent states to derive the classical limit of the two-photon Dicke model in the normal phase. We confirmed the validity of this classical limit by determining the ground-state energy and establishing a quantum-classical correspondence between classical trajectories (via Poincar\'e sections) and quantum spectral statistics. In particular, we identified the onset of classical and quantum chaos in the high-energy regime of the system. Furthermore, we highlighted an intriguing regime of mixed dynamics~\cite{Berry1984,Robnik1998,Robnik2023}, which may offer a fertile ground for studying quantum localization phenomena~\cite{Batistic2010,Batistic2013a,Batistic2013b}. Of special interest is how this mixed classical behavior manifests in the quantum domain, where the coexistence of ordered and disordered structures in the Peres lattice coincides with a mixed spectral distribution [see Figs.~\ref{fig:PeresLattice_gamma0p3_S1}(b1)-\ref{fig:PeresLattice_gamma0p3_S1}(b4), column view].

Future directions include extending the analysis of classical dynamics into the superradiant phase and exploring whether classical and quantum chaos emerge in this regime. Additional avenues of research involve investigating quantum scarring, thermalization, and multifractality in the normal phase of the two-photon Dicke model, drawing parallels with known phenomena in the single-photon Dicke model~\cite{Pilatowsky2021NatCommun,Villasenor2023,Bastarrachea2024}. Finally, it would be intriguing to explore the experimental verification of dynamical signatures of quantum chaos~\cite{Das2025,Vallejo2025} in this system.

\section*{Acknowledgments}
\label{sec:Acknowledgments}

We acknowledge the support of the Computation Center - ICN, in particular to Enrique Palacios, Luciano D\'iaz, and Eduardo Murrieta. F.R., V.S.M-G, D.Y.C, and J.G.H acknowledge partial financial support from the DGAPA-UNAM Project No. IN109523. D.V. was supported by the Slovenian Research and Innovation Agency (ARIS) under Grants No. J1-4387 and No. P1-0306.

\appendix

\section{Mean-field analysis with coherent states}
\label{app:MeanField}

A mean-field approximation is achieved by taking a trial state and computing the expectation value of an algebraic quantum Hamiltonian. The two-photon Dicke model is an algebraic Hamiltonian composed of two Hilbert subspaces, one bosonic $\mathcal{H}_{\text{B}}$ with infinite dimension, and one atomic $\mathcal{H}_{\text{A}}$ with finite dimension $2j+1$. Therefore, we use Glauber and Bloch coherent states of minimum uncertainty~\cite{Arecchi1972,Zhang1990}, given in Eq.~\eqref{eq:GlauberState} and Eq.~\eqref{eq:BlochState}, which are related to the bosonic and atomic subspace, respectively. By taking the expectation value of the two-photon Dicke Hamiltonian in Eq.~\eqref{eq:DickeHamiltonian} under tensor product of the last coherent states in Eq.~\eqref{eq:GlauberBlochState}, we get the expression
\begin{align}
    \label{eq:ClassicalHamiltonian}
    \langle\mathbf{x}|\hat{H}_{\text{D}}|\mathbf{x}\rangle
    = & \omega|\alpha|^{2}-j\omega_{0}\frac{1-|\beta|^{2}}{1+|\beta|^{2}} \\
    & + \gamma\left[(\alpha^{\ast})^{2} + \alpha^{2}\right]\frac{\beta^{\ast} + \beta}{1+|\beta|^{2}} . \nonumber
\end{align}

Each coherent-state parameter $(\alpha,\beta)\in\mathbb{C}$ can be linked to a pair of canonical position and momentum variables in phase space~\cite{Villasenor2024ARXIV}
\begin{align}
    \alpha & = \sqrt{\frac{j}{2}}(q+ip), \label{eq:qp1}\\
    \beta & = \frac{Q+iP}{\sqrt{4-Q^{2}-P^{2}}},
    \label{eq:qp2}
\end{align}
such that, $(q,p)$ identify the bosonic classical coordinates and $(Q,P)$ the atomic classical coordinates. By substituting the coherent-state parameters $\alpha$ and $\beta$ in Eq.~\eqref{eq:ClassicalHamiltonian} and scaling the expectation value by the system size $j$, we get the classical two-photon Dicke Hamiltonian given in Eq.~\eqref{eq:ClassicalDickeHamiltonian}.

The minimization of the Hamilton equations of motion for the classical two-photon Dicke Hamiltonian yields the extreme values $\mathbf{x}_{\text{gs}}=(0,0;0,0)$, which correspond to the normal phase of the system. Evaluating Eq.~\eqref{eq:ClassicalDickeHamiltonian} with these coordinates allows us to determine the ground-state energy for the normal phase
\begin{equation}
    \epsilon_{0}=\frac{E_{0}}{j}=h_{\text{D}}(\mathbf{x}_{\text{gs}})=-\omega_{0} ,
\end{equation}
which is the same expression obtained with the mean-field approximation using squeezed states~\cite{Garbe2017}.

\section{Bogoliubov transformation}
\label{app:Bogoliubov}

The Hamiltonian in Eq.~\eqref{eq:DickeHamiltonian} becomes integrable when setting $\omega_0=0$,
\begin{equation}
    \label{eq:Htilde}
    \hat{H}'_{\text{D}} = \omega \hat{a}^{\dagger}\hat{a} + \frac{2\gamma}{\mathcal{N}} \left( \hat{a}^{\dagger 2} + \hat{a}^{2} \right) \hat{J}_{x},
\end{equation}
where we used the expression $\hat{J}_{x}=(\hat{J}_{+}+\hat{J}_{-})/2$.

The last Hamiltonian commutes with the operator $\hat{J}_{x}$, $[\hat{H}'_{\text{D}},\hat{J}_x]=0$. Both operators share a set of eigenstates $|n; j, m_x\rangle = |n\rangle\otimes|j,m_x\rangle$, such that, $\hat{a}^{\dagger}\hat{a}|n; j,m_x\rangle = n|n; j,m_x\rangle$ and $\hat{J}_x|n; j,m_x\rangle = m_x|n; j,m_x\rangle$. Using the basis $|j,m_x\rangle$, we obtain $2j+1$ Hamiltonians for each eigenvalue $m_x$,
\begin{align}
    \label{eq:Htildemx}
    \hat{H}'_{m_x} = & \langle j, m_x | \hat{H}_{\text{D}} | j, m_x \rangle \\
    = & \omega \hat{a}^{\dagger}\hat{a} + \frac{2\gamma}{\mathcal{N}} \left( \hat{a}^{\dagger 2} + \hat{a}^{2} \right) m_x , \nonumber
\end{align}
which can be diagonalized by employing a Bogoliubov transformation of the form~\cite{Emary2002}
\begin{align}
    \label{eq:aadaggertrans}
    \hat{a} = & \frac{\hat{c} + \sigma \hat{c}^\dagger}{\sqrt{1 - |\sigma|^2}} , \\
    \hat{a}^\dagger = & \frac{\hat{c}^\dagger + \sigma^{\ast}
    \hat{c}}{\sqrt{1 - |\sigma|^2}},
\end{align}
where $\sigma \in \mathbb{C}$ and the operators $\hat{c}$ and $\hat{c}^\dagger$ satisfy the Heisenberg-Weyl algebra with $[\hat{c},\hat{c}^{\dagger}]=\hat{\mathbb{I}}$.

The new Hamiltonian is given by
\begin{equation}
    \label{eq:Htildemx_transf}
    \hat{H}'_{m_x} = A_{m_x} \hat{c}^{\dagger} \hat{c} + B_{m_x} \hat{c}^{\dagger2} +  B_{m_x}^{\ast} \hat{c}^2 + C_{m_x} ,
\end{equation}
where
\begin{align}
    A_{m_x} = & \frac{1}{1 - |\sigma|^2}\left[ \omega\left(1 + |\sigma|^2\right) + \frac{4\gamma m_x}{\mathcal{N}}(\sigma^{\ast}+\sigma) \right] , \\
    B_{m_x} = & \frac{1}{1 - |\sigma|^2}\left[ \omega\sigma + \frac{2\gamma m_x}{\mathcal{N}}\left(1 + \sigma^2\right) \right] , \\
    C_{m_x} = & \frac{1}{1 - |\sigma|^2}\left[ \omega|\sigma|^2 + \frac{2\gamma m_x}{\mathcal{N}}(\sigma + \sigma^{\ast})\right] ,
\end{align}
and it becomes diagonal for the condition 
\begin{equation}
    \label{eq:sigmapm}
   \sigma_{m_x} = \sigma_{m_x}^{\ast} =\frac{-1 \pm \sqrt{1 - 4 \lambda_{m_x}^2}}{2 \lambda_{m_x}}, 
\end{equation}
where $\lambda_{m_x} \equiv 2\gamma m_x / (\omega\mathcal{N})$.

The spectral collapse occurs at $\gamma_{\text{sc}} = \omega / 2$. In this work we work in the region $\gamma < \gamma_{\text{sc}}$, where $\sigma$ is a real quantity. In addition, we study the spectrum characterized by increasing energy values. The last condition is satisfied by selecting the positive root of Eq.~\eqref{eq:sigmapm}, giving a positive coefficient of the operator $\hat{c}^\dagger\hat{c}$. Taking the last considerations, we arrive at the Hamiltonian presented in Eq.~\eqref{eq:Hanalytic}.

\begin{figure}[ht]
\centering
\includegraphics[width=1\columnwidth]{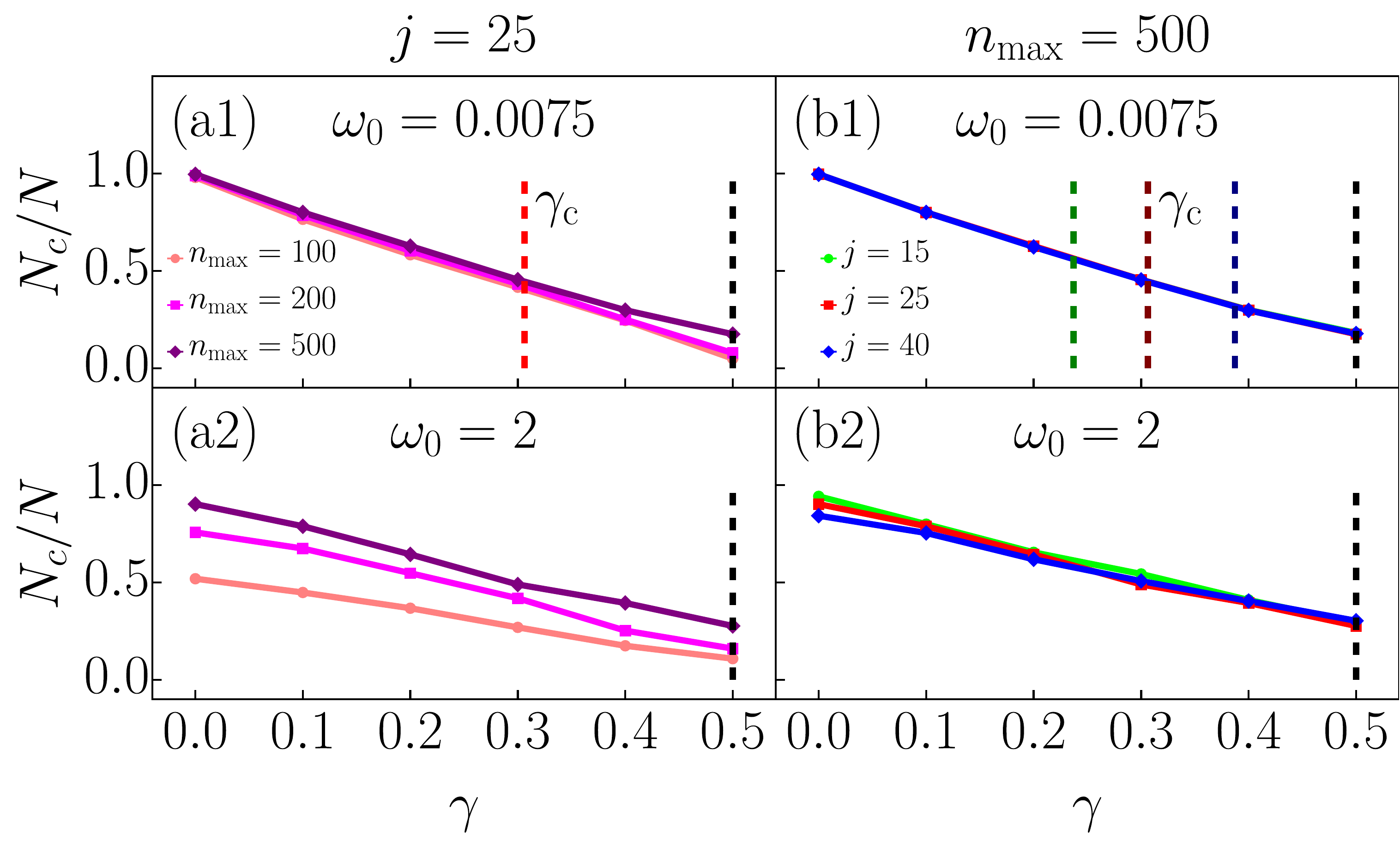}
\caption{Fraction of converged eigenstates $N_{c}/N$ as a function of the coupling strength $\gamma$. The fraction of converged eigenstates was computed using the criterion of Eq.~\eqref{eq:Convergence} with a tolerance parameter $\delta=10^{-3}$. In the first column, we set a system size $j=25$, two values of the atomic level splitting (a1) $\omega_{0}=0.0075$ and (a2) $\omega_{0}=2$, and three truncation values $n_{\max}=100,200$, and 500. In the second column, we set a truncation value $n_{\max}=500$, two values of the atomic level splitting (b1) $\omega_{0}=0.0075$ and (b2) $\omega_{0}=2$, and three system sizes $j=15,25$, and 40. In panel (a1), the vertical red dashed line represents the critical coupling strength $\gamma_{\text{c}}=\sqrt{\omega\omega_{0}j/2}$. In panel (b1), the vertical colored dashed lines represent the critical coupling strength for each system size $j=15,25$, and 40. In all panels, the vertical black dashed line represents the coupling strength of the spectral collapse $\gamma_{\text{sc}}=\omega/2$. The field frequency is $\omega=1$.}
\label{fig:Convergence}
\end{figure}

\section{Numerical diagonalization and convergence}
\label{app:Diagonalization}

The diagonalization of the two-photon Dicke Hamiltonian in Eq.~\eqref{eq:DickeHamiltonian} is performed using a basis formed by the tensor product of Fock states $|n\rangle$ for the bosonic subspace and angular-momentum states $|j,m_{z}\rangle$ for the atomic subspace
\begin{equation}
    \label{eq:FockBasis}
    |n;j,m_z\rangle \equiv |n\rangle\otimes|j,m_z\rangle ,
\end{equation}
where $n=0,1,\dots,\infty$ and $m_{z}=-j,-j+1,\ldots,j-1,j$. The matrix elements of the Hamiltonian in this basis are given by
\begin{gather}
    \langle n';j,m'_{z}|\hat{H}_{\text{D}}|n;j,m_{z}\rangle = \left(\omega n + \omega_{0} m_{z}\right)\delta_{n',n}\delta_{m'_{z},m_{z}} \\
    + \frac{\gamma}{\mathcal{N}}\left( A_{n}\delta_{n',n+2} + A_{n-2}\delta_{n',n-2} \right)B_{m'_{z},m_{z}} , \nonumber
\end{gather}
where $A_{n} = \sqrt{(n+1)(n+2)}$, 
\begin{equation}
    B_{m'_{z},m_{z}} = C_{m_{z}}^{+}\delta_{m'_{z},m_{z}+1} + C_{m_{z}}^{-}\delta_{m'_{z},m_{z}-1},
\end{equation}
and $C_{m_{z}}^{\pm} = \sqrt{j(j+1)-m_{z}(m_{z}\pm1)}$.

For numerical calculations, the Hilbert space is truncated to a finite dimension by imposing a maximum photon number $n_{\max}$. Such truncation can lead to spurious numerical solutions, necessitating a convergence criterion. Here, we define a convergence criterion for the eigenstates of the two-photon Dicke Hamiltonian analogous to that used for the single-photon Dicke Hamiltonian~\cite{Bastarrachea2014PSa,Bastarrachea2014PSb}.

The eigenstates of the two-photon Dicke Hamiltonian satisfy the eigenvalue equation $\hat{H}_{\text{D}}|E_{k}\rangle = E_{k}|E_{k}\rangle$, and can be expanded in the diagonalization basis as
\begin{equation}
    \label{eq:EigenstatesFock}
    |E_{k}\rangle = \sum_{n,m_{z}}c_{n,m_{z}}^{k}|n;j,m_{z}\rangle,
\end{equation}
where $c_{n,m_{z}}^{k}=\langle n;j,m_{z}|E_{k}\rangle$. We define the probability distribution of finding $n$ photons in the $k$th eigenstate
\begin{equation}
    P_{n}^{k} = \sum_{m_{z}}|c_{n,m_{z}}^{k}|^{2},
\end{equation}
where $n=0,1,\ldots,n_{\max}$. By evaluating this distribution at the truncation values $n_{\max}$ and $n_{max}-1$, we impose the condition
\begin{equation}
    \label{eq:Convergence}
    P_{n_{\max}}^{k} + P_{n_{\max}-1}^{k} \leq \delta,
\end{equation}
where $\delta$ is a small tolerance parameter. When this condition is satisfied, it guarantees that the wavefunction of the $k$th. eigenstate is converged within the truncated Hilbert space and will remain unchanged upon further increasing $n_{\max}$. This criterion is checked for each eigenstate, ordered by ascending energy. The lowest-energy eigenstate that violates this inequality determines the set of converged eigenstates (eigenenvalues).

In the following, we illustrate the effects of our convergence method. We first examine how the truncation value $n_{\max}$ impacts the fraction of converged eigenstates (eigenvalues). This fraction is defined as the ratio of the number of converged eigenstates $N_{c}$, identified using the criterion in Eq.~\eqref{eq:Convergence}, to the total number of eigenstates $N$.

In Fig.~\ref{fig:Convergence}(a1), we consider an atomic level splitting of $\omega_{0}=0.0075$, where there is a coexistence of normal and superradiant phases before the spectral collapse. Figure~\ref{fig:Convergence}(a2) examines a value of $\omega_{0}=2$, where only the normal phase exists prior to the spectral collapse. Both scenarios were computed for a fixed system size of $j=25$ and three truncation values $n_{\max}=100,200$, and 500. In Fig.~\ref{fig:Convergence}(a1), we observe that the fraction $N_{c}/N$ increases slowly at the coupling strength corresponding to the spectral collapse $\gamma_{\text{sc}}=\omega/2$, particularly for the largest truncation value $n_{\max}=500$. For the other coupling strengths $\gamma<\omega/2$, the fraction remains relatively constant, regardless of the truncation value. In contrast, as shown in Fig.~\ref{fig:Convergence}(a2), increasing the truncation value consistently enhances the fraction of converged states across all coupling strengths.

Finally, we examine the effect of system size $j$ on the fraction $N_{c}/N$. In Figs.~\ref{fig:Convergence}(b1) and~\ref{fig:Convergence}(b2), we analyze the previous values of atomic level splitting $\omega_{0}=0.0075$ and $\omega_{0}=2$, respectively. Both cases were computed with a fixed truncation value of $n_{\max}=500$ and for three different system sizes $j=15,25$, and 40. The results in Figs.~\ref{fig:Convergence}(b1) and~\ref{fig:Convergence}(b2) indicate that, for a constant truncation value, increasing the system size does not significantly alter the fraction $N_{c}/N$.

\bibliography{main}

\end{document}